\documentclass[iop]{emulateapj}

\usepackage{graphicx}
\usepackage{amsmath}
\usepackage{amssymb}
\usepackage{natbib}
\usepackage{color}
\usepackage{enumitem}

\usepackage{txfonts}
\begin{document}

\title{Radiation Hydrodynamical Turbulence in Protoplanetary Disks:\\ Numerical Models and Observational Constraints}

   \author{Mario Flock\altaffilmark{1,9}, Richard P. Nelson\altaffilmark{2,9}, Neal J. Turner\altaffilmark{1,9}, Gesa H.-M. Bertrang\altaffilmark{3,4}, Carlos Carrasco-Gonz\'{a}lez\altaffilmark{5}, Thomas Henning\altaffilmark{6}, Wladimir Lyra\altaffilmark{7,1} and Richard Teague\altaffilmark{6,8}}
   \affil{$^1$Jet Propulsion Laboratory, California Institute of Technology, Pasadena, California 91109, USA}
   \affil{$^2$Astronomy Unit, Queen Mary University of London, Mile End Road, London E1 4NS, UK}
   \affil{$^3$Universidad de Chile, Departamento de Astronomia, Casilla 36-D, Santiago, Chile}
   \affil{$^4$Millennium Nucleus Protoplanetary Disks in ALMA Early Science, Universidad de Chile, Casilla 36-D, Santiago, Chile}
   \affil{$^5$Instituto de Radioastronom\'{i}a y Astrof\'{i}sica UNAM, Apartado Postal 3-72 (Xangari), 58089 Morelia, Michoac\'{a}n, M\'{e}xico}
   \affil{$^6$Max-Planck-Institut f\"ur Astronomie, K\"onigstuhl 17, D-69117 Heidelberg, Germany}
   \affil{$^7$Department of Physics and Astronomy, California State University Northridge, 18111 Nordhoff St, Northridge, CA 91330, USA}
   \affil{$^8$Department of Astronomy, University of Michigan, 311 West Hall, 1085 S. University Ave, Ann Arbor, MI 48109, USA}
   \affil{$^9$Kavli Institute For Theoretical Physics, University of California, Santa Barbara, CA 93106, USA} 
   \email{mflock@caltech.edu}
   \date{}

  \begin{abstract} 
{Planets are born in protostellar disks, which are now observed with enough resolution to address questions about internal gas flows. Candidates for driving the flows include magnetic forces, but ionization state estimates suggest much of the gas mass decouples from magnetic fields.  Thus, hydrodynamical instabilities could play a major role.  We investigate disk dynamics under conditions typical for a T Tauri system, using global 3D radiation hydrodynamics simulations with embedded particles and a resolution of 70 cells per scale height.  Stellar irradiation heating is included with realistic dust opacities.  The disk starts in joint radiative balance and hydrostatic equilibrium.  The vertical shear instability (VSI) develops into turbulence that persists up to at least 1600 inner orbits (143 outer orbits).  Turbulent speeds are a few percent of the local sound speed at the midplane, increasing to 20\%, or 100 m/s, in the corona.  These are consistent with recent upper limits on turbulent speeds from optically thin and thick molecular line observations of TW Hya and HD 163296.  The predominantly vertical motions induced by the VSI efficiently lift particles upwards.  Grains 0.1 and 1 mm in size achieve scale heights greater than expected in isotropic turbulence.  We conclude that while kinematic constraints from molecular line emission do not directly discriminate between magnetic and nonmagnetic disk models, the small dust scale heights measured in HL Tau and HD 163296 favor turbulent magnetic models, which reach lower ratios of the vertical kinetic energy density to the accretion stress.}
\end{abstract}


\shorttitle{Radiation hydrodynamic turbulence in protoplanetary disks}
\shortauthors{Flock et al.}

\maketitle

\section{Introduction}
Understanding the transport of angular momentum and the underlying dynamics in protoplanetary disks is one of the key questions in planet and star formation research. Since the classical $\alpha$-disk theory presented by \citet{sha73}, in which the transport is described by an effective turbulent viscosity, there has been an ongoing search for a mechanism capable of generating turbulence in accretion disks, as pure molecular viscosity is inefficient in transporting angular momentum \citep{pri81}. A powerful mechanism providing turbulence is the magneto-rotational instability \citep{bal91,bal98}. However, recent theoretical studies have shown that magnetically driven instabilities can be suppressed in some regions of the disk \citep{tur14,arm15} due to the different non-ideal terms of Ohmic resistivity, ambipolar diffusion and Hall drift \citep{war07,sal08,bai11,bai14,dzy13}. In such regions, hydrodynamical instabilities could play an important role. Examples include the Goldreich$-$Schubert$-$Fricke (GSF) instability \citep{gol67,fri68}, the convective instability \citep{cam73}, the Papaloizou$-$Pringle instability \citep{pap84}, the baroclinic instability \citep{kla03,les10}, the Loren-Bate instability \citep{lor15}, the convective overstability \citep{lyr14,kla14}, the zombie vortex instability \citep{mar13,mar15,les16,umu16b} or the spiral wave instability \citep{bae16} which could appear in self-gravitating disks or in disks perturbed by a massive planet \citep{poh15}. %
The GSF, first considered as acting on stellar interiors, is now of interest for accretion disks \citep{urp98,urp03,arl04}. In particular, it may operate in protostellar disks' outer regions, where thermal relaxation is faster than the orbital rotation \citep{nel13,lin15}. In this context, the instability is known as the vertical shear instability, or VSI. Hydrodynamical calculations indicate the VSI produces stress-to-pressure ratios, $\alpha$, ranging from $\sim 10^{-3}$ with an isothermal equation of state \citep{nel13} down to $10^{-4}$ when considering detailed radiative heating and cooling \citep{sto16}.
%

Recent high resolution observations of protoplanetary disks at millimeter wavelengths have revealed astonishing sub-structures in the thermal dust emission \citep{par15,car16,and16,ise16,fed17,mac17}. For the first time, the disk scale height is resolved in to 10 AU. Detailed radiative transfer models of these systems \citep{pin16,vanb17,liu17} support a small stress-to-pressure ratio $\alpha \ll 0.01$, based on best-fit models of the dust disk scale height in the disk system HL Tau. An opportunity to obtain observational constraints on the magnetic field in disks could be the measurements of the linearly polarized continuum and line emission. The recent ALMA observations of HD 142527 present polarization maps at 0.874 mm \citep{kat16} with 70 AU resolution. Those maps show a clear radial symmetry structure inside a radius of 150 AU with a strength of the polarization degree between 0.1 to 14 \%. Both the radial pattern and the degree of the polarization of this observation match very well with the results of thermal dust polarization radiation transfer models from 3D non-ideal MHD models \citep{ber17b,ber17} based on the simulation results of \citep{flo15}. However, also dust scattering plays an important role \citep{kat15,kat17}, especially in regions where the radiation field becomes an-isotropic, which happens close to the optical depth $\tau=1$ transition and where the radiation field can induce grain alignment \citep{taz17}. A direct evidence for magnetic fields in disks based on these constraints is still missing. 

The kinematics in disks have also been measured directly. One of the first indications of subsonic turbulence in protoplanetary disks was found by the SMA and PdBI telescopes for the systems TW Hya and HD 163296 \citep{hug11} and for DM Tau \citep{gui12}. 
New observational constraints on the disk kinematics with ALMA confirm subsonic turbulent velocities in the disk \citep{fla15,tea16,fla17}. 
In this work we present new high-resolution global radiation hydrodynamic models of the outer protoplanetary disk. Together with our previous magnetic disk models \citep{flo15} we will compare both sets of simulation results with the recently obtained observational constraints and determine which model fits best. 
The new radiation hydrodynamic stratified disk simulations include irradiation by the star, realistic dust opacities and achieve a resolution of over 70 cells per disk scale height. 
Compared to previous models, this work achieves three main advances: (1) having realistic initial conditions which are in full radiation hydrostatic equilibrium, (2) radiation hydrodynamic simulations with realistic heating and cooling, and (3) high-resolution with a high order numerical scheme. 

In this work we compare high-resolution global radiation hydrodynamic models with these recently obtained observational constraints. 

The structure of the paper is as follows: In Section 2 we present the method, the equations and the initial conditions in hydrostatic equilibrium. In Section 3 we present the simulation results. In Section 4 we compare the model with recent observational constraints. Section 5 and 6 follows with the discussion and the conclusion. 

\section{Numerical method and disk model}
The radiation hydrodynamic equations are solved using the hybrid flux limited diffusion and irradiation method developed by \citet{flo13} as implemented in the current version 4.2 of the PLUTO code \citep{mig07}. For this work we choose the high-order piece-wise parabolic method (PPM), the Harten-Lax-Van Leer approximate Riemann Solver with the contact discontinuity (HLLC), the FARGO scheme \citep{mas00,mig12a} and the Runge-Kutta time integrator. The Courant number is set to 0.3. This configuration provides very low intrinsic numerical dissipation which is favored for simulating hydrodynamic instabilities. The equations solved are: 
\begin{eqnarray}
 \frac{\partial \mathrm{\rho}}{\partial \mathrm{t} } + \nabla \cdot \left [  \mathrm{\rho} \vec{v}\right ] &=&
0 \, , \label{eq:MDH_RHO} \\
 \frac{\partial \mathrm{\rho} \vec{v}}{\partial \mathrm{t}} + \nabla \cdot \left [
  \mathrm{\rho} \vec{v} \vec{v}^T \right ] + \rm \nabla \mathrm{P}
&=& - \mathrm{\rho} \nabla \mathrm{\Phi} \, , \label{eq:MDH_MOM} \\
 \frac{\partial \mathrm{E}}{\partial \mathrm{t}} + \nabla \cdot \left [ (\mathrm{E} + \mathrm{P})\vec{v} \right ]  &=& - \mathrm{\rho} \vec{v} \cdot \nabla \mathrm{\Phi} \nonumber \\& &  - \kappa_\mathrm{P} \mathrm{\rho} \mathrm{c} (\rm \mathrm{a_R} \mathrm{T}^4 - \mathrm{E_R} )\nonumber\\ & & - \nabla \cdot \mathrm{F}_* \, , \label{eq:MDH_EN} \\
\frac{\partial \mathrm{E_R}}{\partial \mathrm{t}} - \nabla \frac{ \mathrm{c} \mathrm{\lambda}}{ \kappa_\mathrm{R}(\mathrm{T}) \rho} \nabla \mathrm{E_R} &=& \kappa_\mathrm{P}(\mathrm{T}) \mathrm{\rho} \mathrm{c} ( \mathrm{a_R} \mathrm{T}^4 - \mathrm{E_R}) \, , \label{eq:ER} \\
\end{eqnarray}
with the density $\mathrm{\rho}$, the velocity vector $\vec{v}$, the gas pressure 
\begin{equation}
\mathrm{P}= \frac{\mathrm{\rho} \mathrm{k_B} \mathrm{T}} { \mathrm{\mu_g} \mathrm{u}},
\end{equation}
with the gas temperature $\mathrm{T}$, the mean molecular weight $\mathrm{\mu_g}$, the Boltzmann constant $\mathrm{k_B}$, the atomic mass unit $\mathrm{u}$, the total energy $\mathrm{E}= \mathrm{\rho} \mathrm{\epsilon} + 0.5 \mathrm{\rho} \vec{v}^{\,2}$  and the gas internal energy per unit volume $\mathrm{\rho} \mathrm{\epsilon}$. The closure relation between gas pressure and internal energy is provided by $\mathrm{P}=(\mathrm{\Gamma} -1) \mathrm{\rho} \mathrm{\epsilon}$ with the adiabatic index $\mathrm{\Gamma}$. Other symbols include the radiation energy E$_\mathrm{R}$, the irradiation flux $\mathrm{F}_*$, the Rosseland and
Planck opacities $\mathrm{\sigma_R}$ and $\mathrm{\sigma_P}$, the radiation constant $\mathrm{a_R}=4 \mathrm{\sigma_b}/\mathrm{c}$, the Stefan-Boltzmann constant $\mathrm{\sigma_b}$, and c the speed of light. The flux limiter 
\begin{equation}
\mathrm{\lambda} = \frac{2 + \mathrm{R}}{6 + 3\mathrm{R} + \mathrm{R}^2}
\end{equation}
is taken from \citet[][ Eq.~28 therein]{lev81} with 
\begin{equation}
\mathrm{R} = \frac{|\nabla \mathrm{E_R} |}{\mathrm{\sigma_R} \mathrm{E_R}}.
\end{equation}
The gas is a mixture of molecular hydrogen and helium with solar abundance \citep{dec78} so that $\mathrm{\mu_g} =2.35$ and $\mathrm{\Gamma}=1.42$. In this work we consider the frequency integrated irradiation flux at a radial distance $\mathrm{r}$ to be
\begin{equation}
\rm F_*(r) = \left (  \frac{R_*}{r}\right )^2  \sigma_b T_*^4 e^{-\tau}, 
\label{eq:IRRAD}
\end{equation}
with $\rm T_*$ and $\rm R_*$ being the surface temperature
and radius of the star. The radial optical depth for each meridional height $\theta$ of the
irradiation flux is given by:
\begin{equation}
\rm \tau_*(r,\theta)=\int_{R_*}^r \kappa(T_*) \rho_{dust}(r,\theta)  dr = \tau_0 + \int_{R_0}^r\kappa(T_*) \rho_{dust}(r,\theta)  dr \, ,
\label{eq:TAU}
\end{equation}
where $\rm R_0$ denotes the inner radius of the computational domain. The quantity $\rm \tau_0$ is
the inner optical depth provided by material located between the surface of the star and
$\rm R_0$ with $\rm \tau_{0}=\kappa_* \rho_{R_0} (R_0-3R_*)$. 

We define two opacities, one to the starlight $\kappa_*=1300$~cm$^2$~g$^{-1}$, and the other to the disk's thermal re-emission $\kappa_d=400$~cm$^2$~g$^{-1}$. These are Planck-weighted averages of the frequency-dependent dust opacity, at respectively the stellar temperature and a typical re-emission color temperature of 300~K. The opacities are measured in cross-section per unit mass of dust, and are listed in table~\ref{tab:model}.
%
%
\subsection{Boundary condition and buffer zone}
For all models we use a modified outflow boundary condition, which enforces zero inflow at the radial and $\theta$ boundaries. In addition, we extrapolate the logarithmic density along the meridional direction into the ghost zones. In the radial direction we apply buffer zones: a surface density relaxation from 20-22 AU to the initial profile and a radial velocity damping from 20-21 AU and from 97-100 AU to zero velocity. In the absence of these buffers the emptying of the disk would affect the irradiation layer and hence the temperature in the disk. These buffer zones are excluded when we present our kinematic analysis of the results.
  
\begin{figure}
\resizebox{\hsize}{!}{\includegraphics{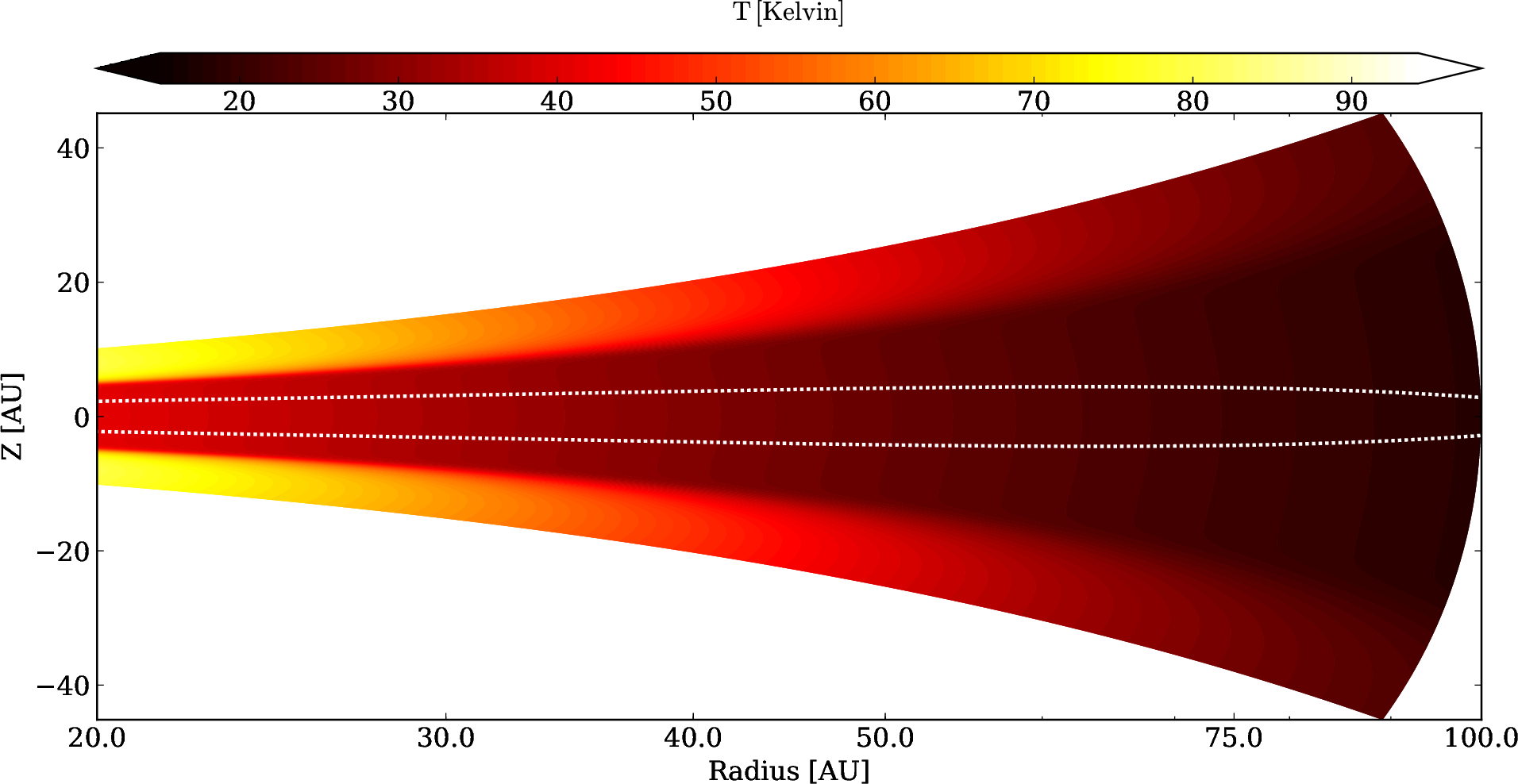}}
\caption{Initial 2D temperature profiles in the R-Z plane for model \texttt{M3}. The temperature rise at the surface marks the surface of unit starlight optical depth. The dotted line shows the surface of vertically integrated optical depth unity for the thermal emission.}
\label{fig:density}
\end{figure}

\subsection{Initial conditions}
For the initial condition, we follow the approach by \citet{flo13} and construct a set of 2D axisymmetric profiles of density, temperature and rotation velocity which are in radiation hydrostatic equilibrium. For the star and disk parameters, we use a model which has been applied to explain various spatially-resolved multi-band observations of protostellar disks \citep{wol03,wol08,sch08,sch09,sau09,mad12,liu12,gra13}. The model uses parameters of a typical T Tauri system. For example, similar density and temperature profiles are found in the TW Hya disk, as discussed in section 4. The star and disk parameters are summarized in Table~\ref{tab:model}. We assume a total dust-to-gas mass ratio of $10^{-2}$. Furthermore, we consider for the dust-to-gas mass ratio of small grains ($a=0.1 \mu m$ to $10 \mu m$), relevant for the disk cooling (and thermal opacity), to be $10^{-3}$ and $10^{-4}$, which is reflected in the model names. Fig.~\ref{fig:density} shows the initial temperature profile for the reference model \texttt{M3}. We note again that the dust-to-gas mass ratio in the model name is relevant in determining the opacity of the thermal emission. We assume the dust mass of small grains, relevant for the infrared emission, has been reduced due to dust growth \citep{bir12,flo13}. 
The dust-to-gas mass ratio directly controls the local opacity and so the thermal equilibrium timescale. Model \texttt{M3} presents a 10 times higher dust-to-gas mass ratio for the small grains, compared to model \texttt{M4}. However both models display short thermal equilibration timescales required for the VSI to operate  \citep{lin15}. 

\subsection{Thermal timescale and critical timescale for the VSI}
We calculate the detailed cooling timescales for both initial models and compare them with the critical timescale required for the VSI to operate. The maximum thermal equilibration timescale consistent with VSI is
\begin{equation}
t_c= \frac{h |q|}{\gamma -1} 
\end{equation}
\citep{lin15}, where $q$ is the radial temperature profile exponent and $ h = H/R$ being the ratio of the disk scale height to radius. 
%
%
%
The thermal relaxation timescale can be estimated by looking at the timescales in the optically thin and thick limits. In the optically thin regime, the corresponding length-scale of the radiation transport is given by the mean free path of the photons 
\begin{equation}
l_{thin}= \frac{1}{\kappa_{dg} \rho} 
\end{equation}
which is typically given by the available dust density $\rm \rho_d$ and dust opacity $\kappa_{d}$ per gram of gas mass $\kappa_{dg}=\kappa_d \frac{\rho_d}{\rho}$. 

The radiation diffusion coefficient can be written as 
\begin{equation}
\rm D_{rad}= \frac{16 \sigma_b T^3}{3 \kappa_{dg} \rho^2 C_v} 
\end{equation}
with the specific heat capacity at constant volume $\rm C_v$.

Following \citet{lin15}, the thermal relaxation timescale can be approximated as the sum of the relaxation timescales in the optical thin and thick regime with 
\begin{equation}
t_{relax}=t_{thin}+t_{thick} = \frac{l_{thin}^2}{3 D_{rad}} + \frac{H^2}{D_{rad}}
\end{equation}
using the disk scale height H as the typical perturbation length-scale. 

Using this the optically thin and thick relaxation timescales, normalized by $\Omega_K$, are given by 
\begin{equation}
\tau_{relax}=\tau_{thin}+\tau_{thick} = \frac{C_v \Omega_K}{16 \sigma_b T^3} \left [ \frac{1}{\kappa_{dg}} + 3 H^2 \rho^2 \kappa_{dg}  \right ].
\end{equation}

\begin{figure}
\resizebox{\hsize}{!}{\includegraphics{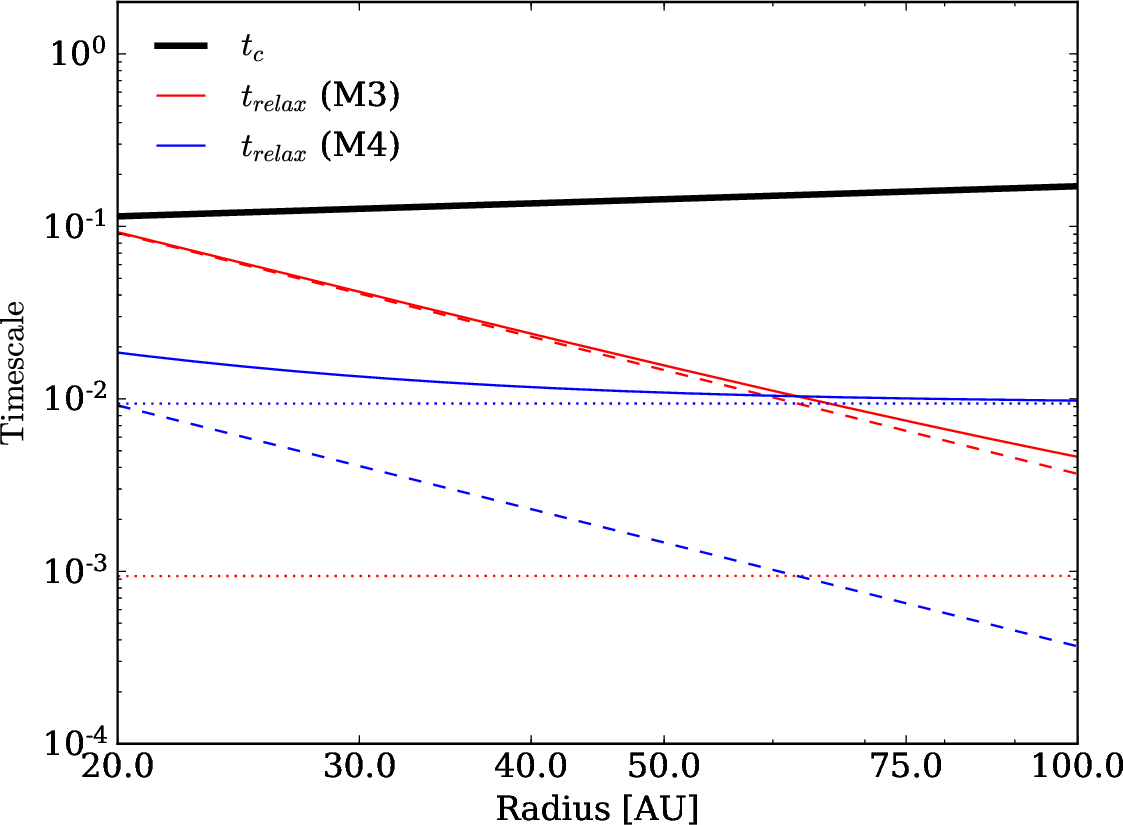}}
\resizebox{\hsize}{!}{\includegraphics{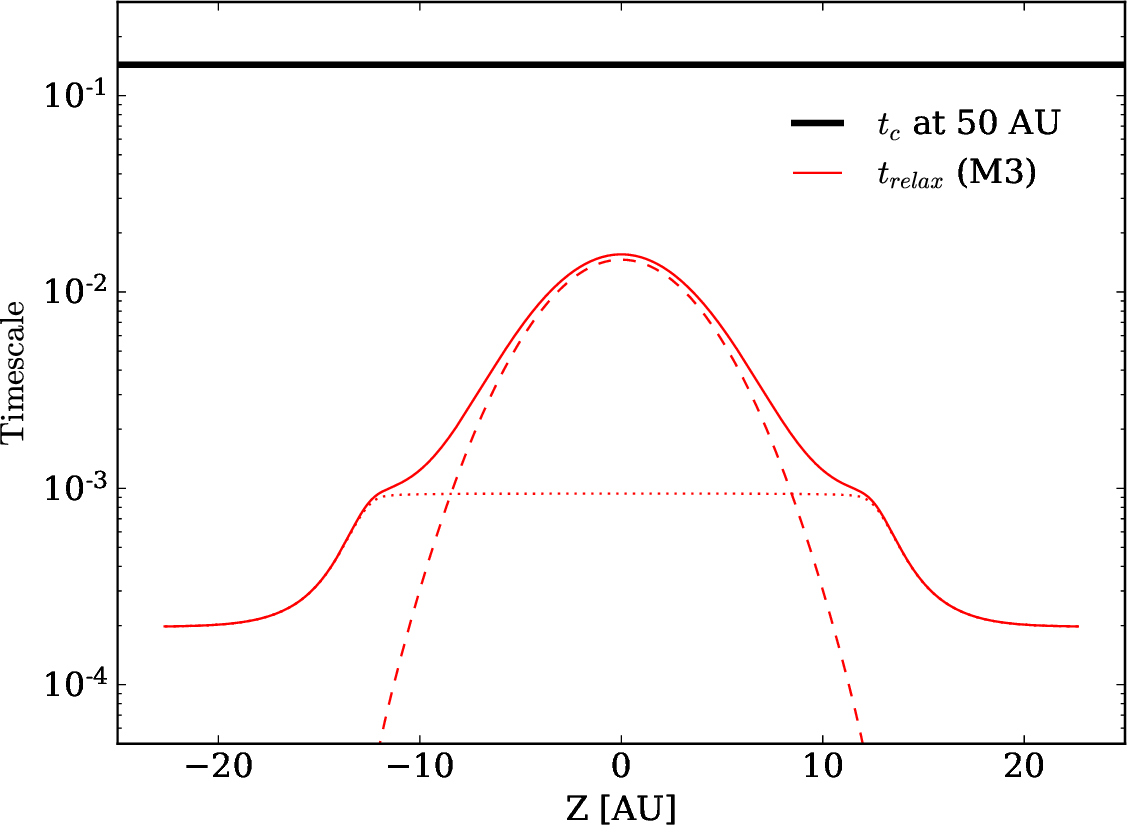}}
\caption{{Top: Radial profiles of the critical timescale for the VSI (black) and the relaxation timescales at the midplane in models M3 (red) and M4 (blue).  All are in units of the local inverse Keplerian frequency. Bottom: Relaxation timescale vs. height at 50 AU in model M3.  Both panels also show the optically thin (dotted line) and optically thick (dashed line) limits.}}
\label{fig:tc}
\end{figure}

Fig.~\ref{fig:tc} compares the critical timescale, $\rm t_c$, for the VSI with the disk thermal timescales for our model. 
The plot shows that the thermal equilibration timescales are lower than the critical timescale and the VSI is expected to operate in the full domain. We note that for model \texttt{M3}, the dominant timescale at the midplane is given by the radiation diffusion while for model \texttt{M4}, the dominant timescale is given in the optical thin limit. We note that local enrichment and depletion of dust grains could lead to variations of the thermal timescales.  

These considerations mean the VSI could in principle operate in model M3 down to 19~AU, a little closer to the star than the domain's innermost radius of 20~AU.
%
This agrees well with work by \citet{mal17}, where the zone of VSI operation based on the radiative relaxation timescale lies between 15 and 180 AU for disk density and temperature profiles similar to our model M3. For model \texttt{M4} the unstable region extents to a radius inwards of 7 AU. This shows that dust depleted atmospheres in proto-planetary disk could support the conditions of the VSI to operate over a large radial extent. In addition, we expect the upper layers of the disk to have conditions conducive to the VSI (noting, however, that we neglect the possible stabilizing effects of magnetic fields in this work).

\begin{table}
\begin{tabular}{lll}
\hline
Surface density & $\rm \Sigma=6.0 \left ( \frac{r}{100 AU} \right )^{-1} g/cm^2$ \\
Stellar parameters & $\rm T_*=4000\, K$, $\rm R_*=2.0\, R_\sun,\, M_*=0.5\, M_\sun$\\
Opacities & $\rm \kappa_*=1300\, cm^2/g$\\
        & $\rm \kappa_d = 400\, cm^2/g$\\ 
\end{tabular}
\caption{Setup parameters for the 3D radiation HD disk models, including the gas surface density, the stellar parameters, the opacities for stellar irradiation and thermal emission.}
\label{tab:model}
\end{table}

%
\begin{table*}[ht]
\begin{tabular}{llllll}
model name & $\rho_{dust}/\rho_{gas}$ & domain size: $\rm R_{in}-R_{out} : Z/R : \Phi_{max}$ & resolution ($N_r \times N_\theta \times N_\phi$) & $\overline{\alpha}$ & runtime [orbits at 20 AU] \\
\hline
\texttt{M3} & $10^{-3}$ & $\rm 20-100 AU\, : \, \pm 0.35 : \, 0.392 rad$ $(22.5^o)$ & $1024 \times 512 \times 256$ & $ 3.7 (\pm 0.6) \times 10^{-5}$ & 470 \\
\texttt{M4} & $10^{-4}$ & $\rm 20-100 AU\, : \, \pm 0.35 : \, 0.392 rad$ $(22.5^o)$ & $1024 \times 512 \times 256$ & $ 3.8 (\pm 1.2) \times 10^{-5}$ & 1610 \\
\end{tabular}
\caption{List of simulations and model parameter. From left to right: the model name, the dust-to-gas mass ratio, the domain size, the resolution, the space and time averaged stress-to-pressure ratio and the total simulation runtime.}
\label{tab:runs}
\end{table*}

\begin{figure}
  \resizebox{\hsize}{!}{\includegraphics{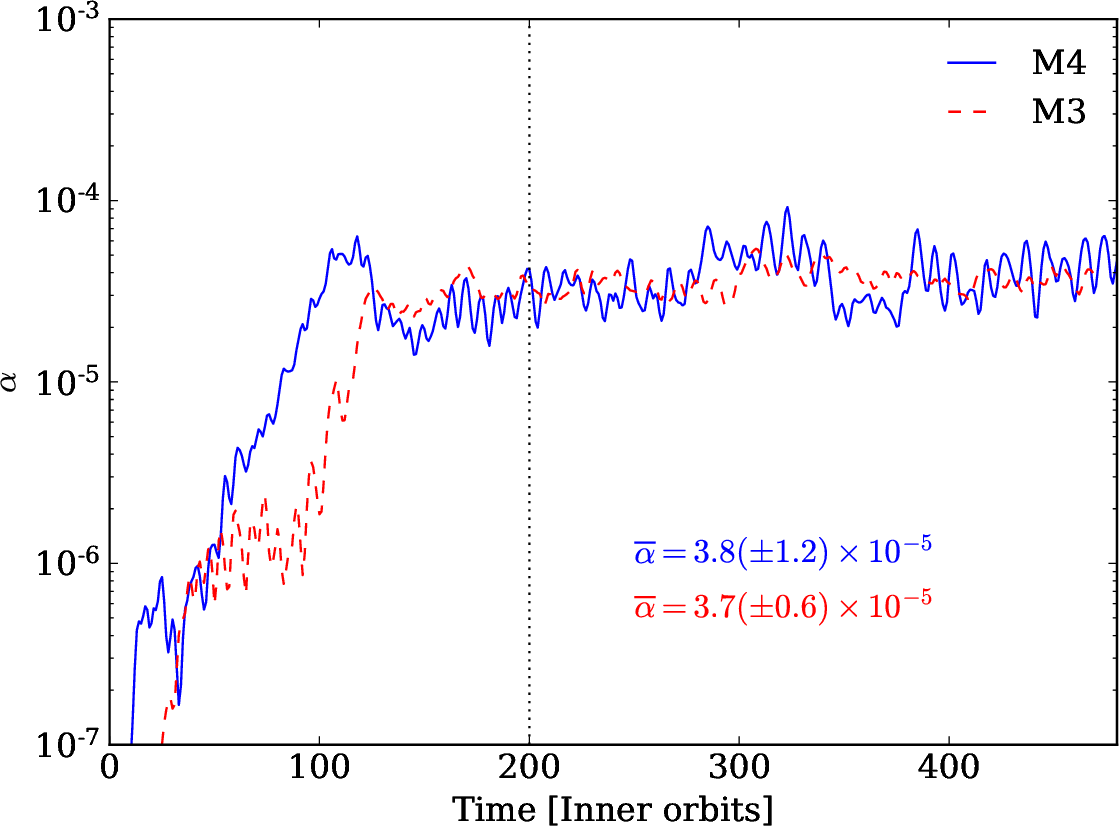}}
\caption{{Time evolution of the stress-to-pressure ratio $\alpha$ in the two models. Also listed are time averages from 200 inner orbits up through the end of each calculation.}} 
\label{fig:alp}
\end{figure}

\section{Simulation results and comparison with observations}
In this section we present the results on the disk dynamics and the turbulent structure. We first investigate the effect of changing the dust-to-gas mass ratio and hence the local opacity, using \texttt{M3} and \texttt{M4}. In Section~\ref{sec:alp} we investigate the turbulent characteristics of the two models. In Section~\ref{sec:line} we summarize recent constraints from measurements of molecular line broadening. In Section~\ref{sec:dustscale} we analyze how grains of representative sizes move in the two models, we determine their dust scale heights from our simulations and compare them to recent best-fit models of HL Tau by \citet{pin16}.\\ 

\subsection{Stress-to-Pressure Ratio}  
\label{sec:alp}
We determine the strength of the turbulence by calculating the stress-to-pressure ratio $\alpha$. For these radiation hydrodynamic models the Reynolds stress $\rm T_{r \phi} $ is the relevant stress which enables transport of angular momentum. We find a typical value by volume-averaging the mass-weighted stress-to-pressure ratio, using
\begin{equation}
\rm \alpha = \frac{ \int \rho \Bigg( \frac{T_{r \phi}}{P} \Bigg)dV} {\int \rho dV} = \frac{ \int \rho \Bigg(
  \frac{\rho v'_{\phi}v_{r}}{P} \Bigg)dV}
    {\int \rho dV}. 
\label{eq:ALPHA}
\end{equation}
We choose mass-weighting for direct comparison with measurements from our magnetohydrodynamical disk models. We compare this approach with a simple ratio of volume-averaged stress to volume-averaged pressure, as in \citet{nel13}, in Appendix A. The two approaches deliver similar results.

The time evolution of the stress-to-pressure ratio $\alpha$ is presented in Fig.~\ref{fig:alp} for both models. After around 100 inner orbits, the value of $\alpha$ saturates and both models show a steady turbulent evolution. Both models show nearly the same saturated value of $\alpha$. For model \texttt{M3} we determine a time averaged value of 3.7 $\times 10^{-5}$, compared to 3.8 $\times 10^{-5}$ for model \texttt{M4}. The time average is done between 200 inner orbits until the end of the simulation run. We note that the fluctuations in the stress-to-pressure ratio around the mean remain small and at a level of 30\%. Table~\ref{tab:runs} summarizes the results.  

\begin{figure*}[ht]
  \resizebox{0.49 \hsize}{!}{\includegraphics{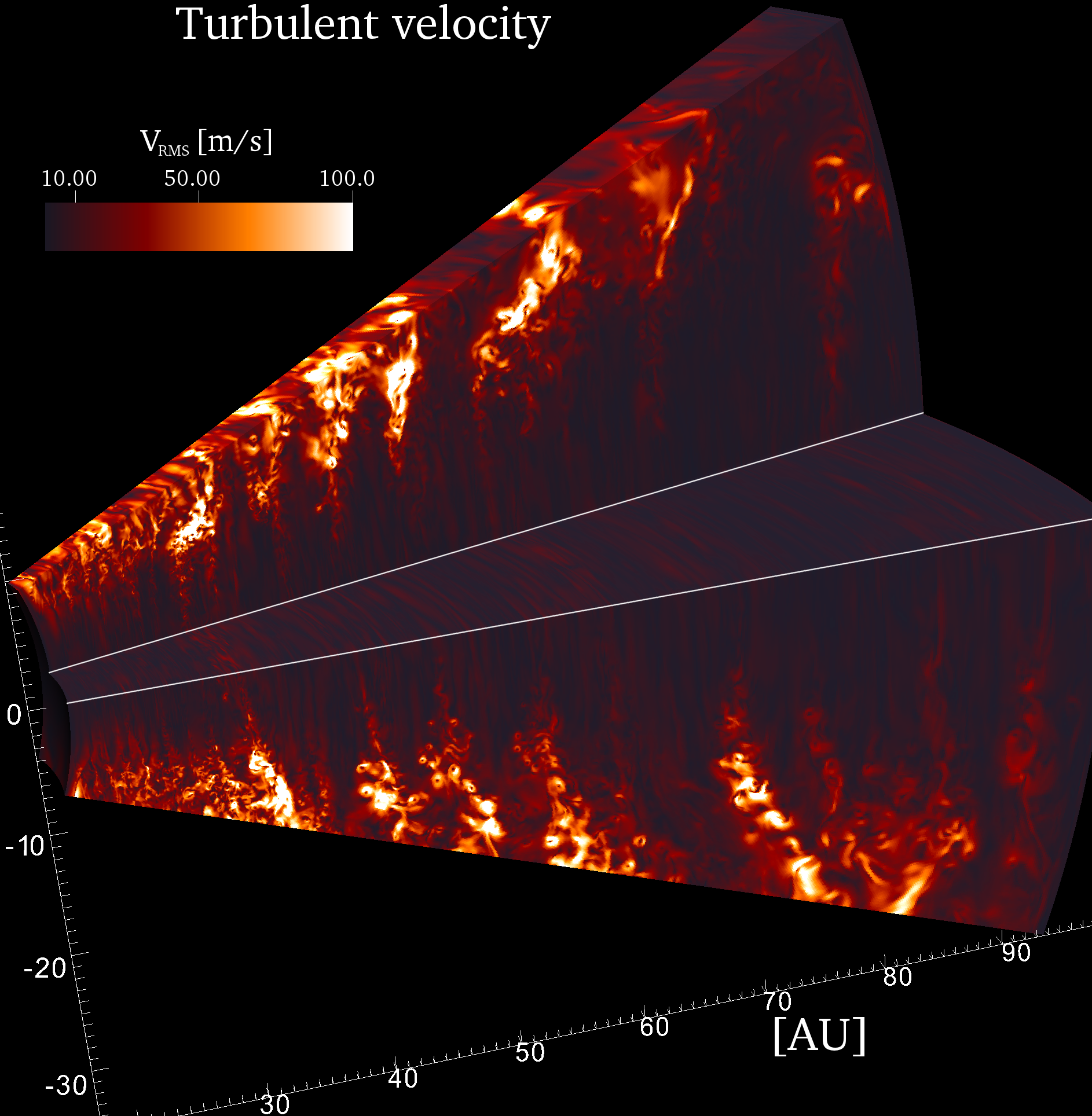}}
  \resizebox{0.49 \hsize}{!}{\includegraphics{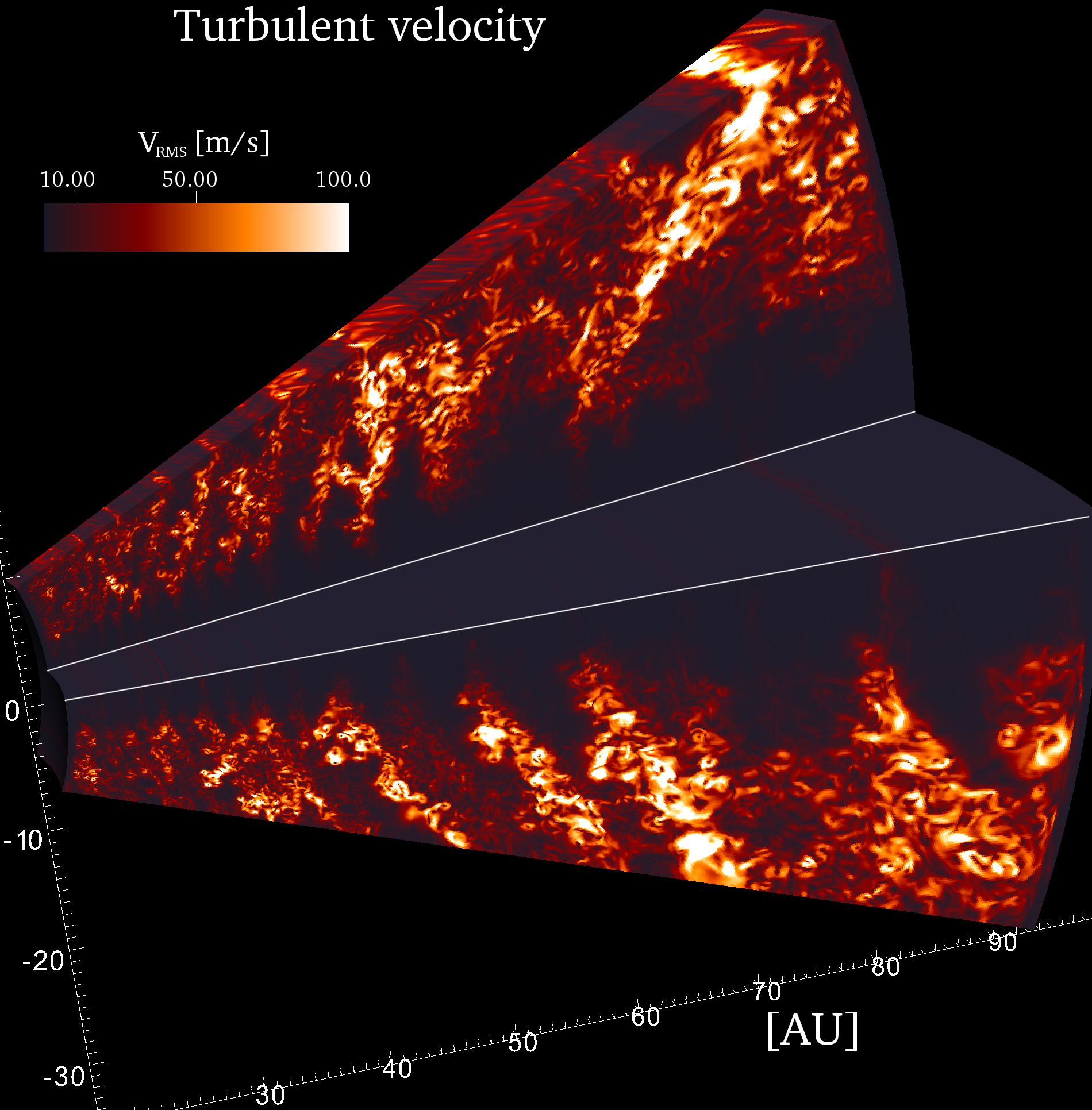}}
\caption{{3D rendering of turbulent speeds in model M3 after 440 inner orbits (left) and model M4 after 1000 inner orbits (right).  In both cases, we see the near side of the domain's lower half, and the far side of its upper half.  The rest of the upper half is cut away to reveal values on the midplane. The turbulent velocity is 10 m/s in the midplane, but exceeds 100 m/s at some points in the corona.}}
\label{fig:vrms3d}
\end{figure*}

Results of a long term evolution are presented in Appendix A, performed for model \texttt{M4} for over 1600 inner orbits. We note that there is a very stable and steady turbulent evolution. Snapshots of the turbulent velocity field in steady state are shown in Fig.~\ref{fig:vrms3d} for both models. The 3D rendering shows the characteristic turbulent structure of the VSI, showing enhanced turbulent motions with a velocity dispersion up to 100 m/s in the disk corona. At the midplane the turbulent velocities are around 10 m/s and below. We emphasize again that model \texttt{M3} and model \texttt{M4} show very similar turbulent behavior. The largest differences which we observe are that the turbulent velocities at the midplane are slighty reduced compared to model \texttt{M3}. This can be already seen by looking at the snapshots of Fig.~\ref{fig:vrms3d}. Further, in model \texttt{M4} we observe an increase of turbulent velocity slightly earlier with height. These differences might be connected to the different temperature profile as in model \texttt{M4} the irradiation heating zone reaches much deeper into the disk due to the lower optical depth. In the following analysis we will focus on model \texttt{M3}.

\subsection{Kinematic Constraints From Line Observations}  
\label{sec:line}
In this section we review recent observational constraints and compare them with the results from our radiation hydrodynamic simulations. 

\begin{figure*}[ht]
  \resizebox{0.49\hsize}{!}{\includegraphics{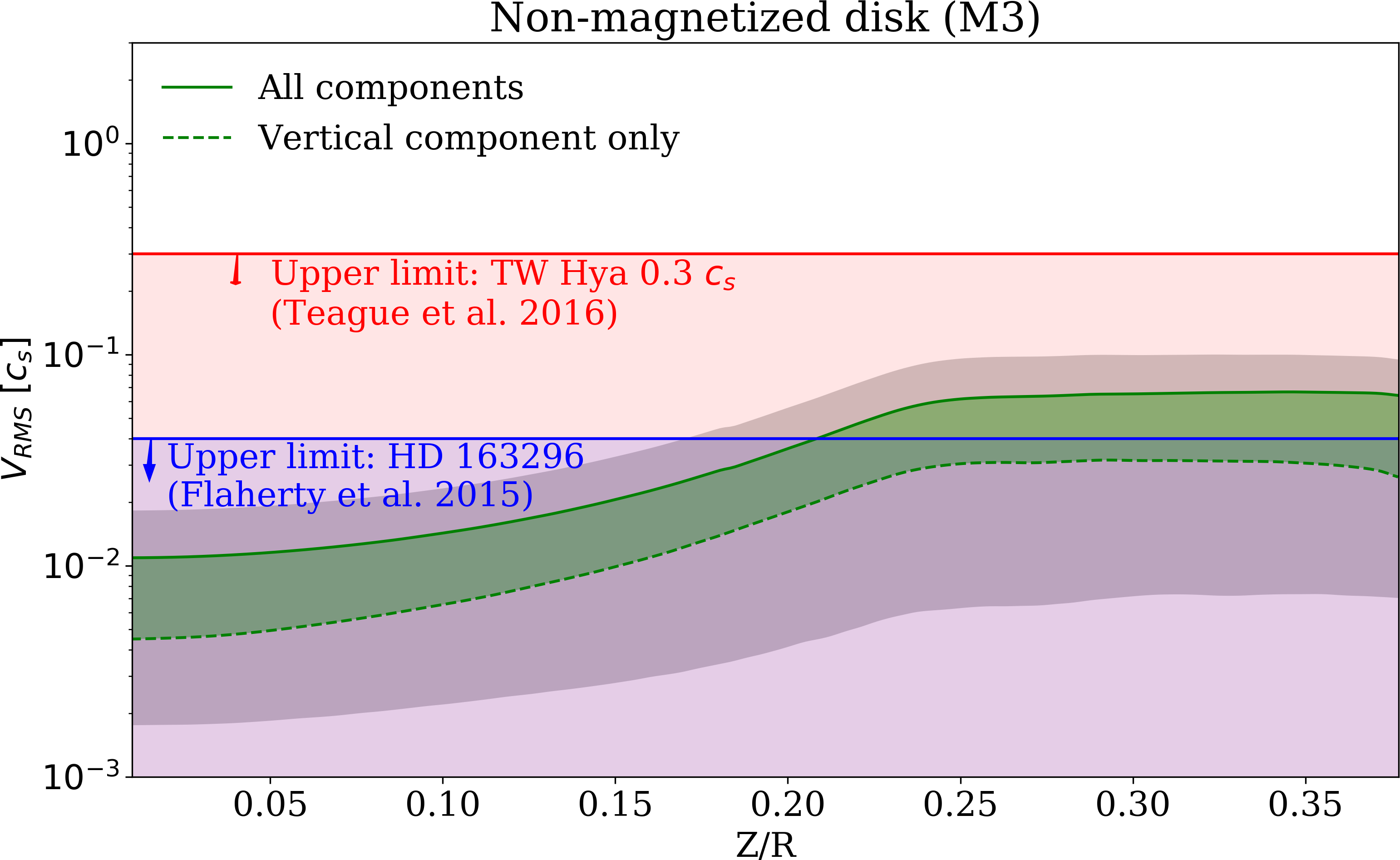}}
  \resizebox{0.49\hsize}{!}{\includegraphics{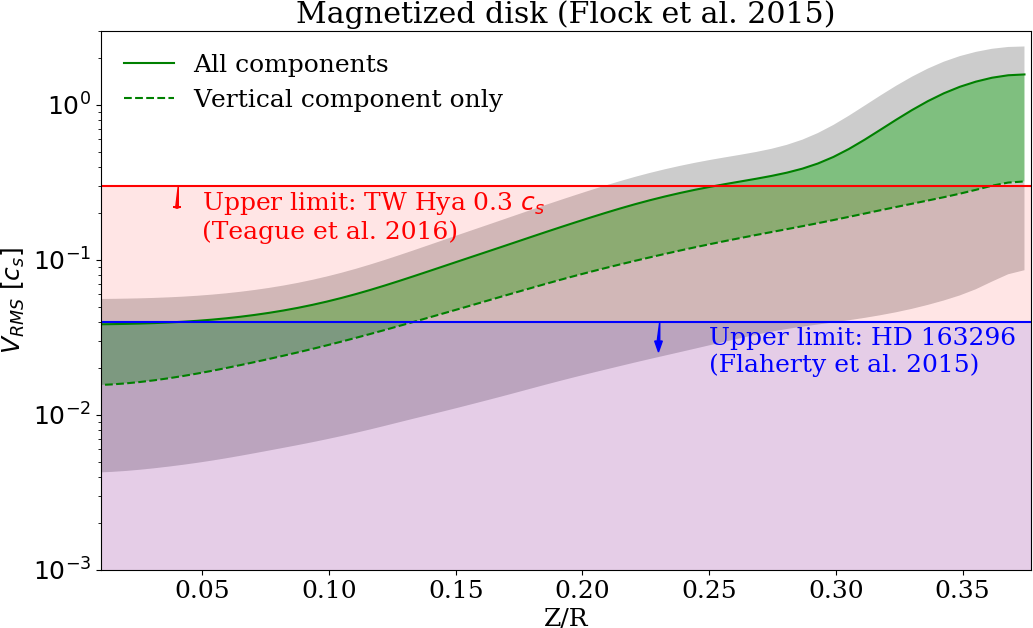}}
\caption{Left: Time and space averaged profile of the turbulent velocity over height for model \texttt{M3}. Right: Time and space averaged profile of the turbulent velocity over height for magnetized disk model by \citet{flo15}. Both profiles show the values at 50 AU. Annotated are the best fit constraints from the systems TW Hya \citep{tea16} and HD 163296 \citep{fla15}. {The green area presents the range of turbulent velocity, for the vertical component only (dashed line) and for all velocity components (solid line). The grey area shows the standard deviation for each component over horizontal planes, averaged in time.}} 
\label{fig:vrms_point}
\end{figure*}

In Fig.~\ref{fig:vrms_point}, left panel, we summarize the model results and compare them with the observational constraints. 
The plot shows the time and spatially averaged turbulent velocity profile over height for model \texttt{M3}. The turbulent velocity in the $r -\theta$ plane is measured by using
\begin{equation}
\left < \overline{v_{RMS}(r,\theta)} \right >= \left < \overline{\sqrt{ (v_r-\overline{v_r})^2+(v_\theta-\overline{v_\theta})^2+(v_\phi-\overline{v_\phi})^2}} \right >, 
\end{equation}
with the overbarred quantities representing the azimuthal average and the symbols $< >$ representing the time average. 
In addition, we calculate the turbulent velocity of the $v_\theta$ component only by using  $ \left < \overline{v^\theta_{RMS}(r,\theta)} \right >= \left < \overline{\sqrt{ (v_\theta-\overline{v_\theta})^2}} \right >$. This case is useful to compare with face-on disk observations in which one is sensitive for only the vertical velocity component. The same approach is used to calculate the standard deviation. For each time output we calculate the standard deviation in the $(r,\theta)$ plane along the $\phi$ direction. In a second step we perform the time average of this spatial standard deviation. The results are presented at the radius of 50 AU using a radial average of $\pm 1$ disk scale height. Fig.~\ref{fig:vrms_point}, left panel, shows that the turbulent velocity at the midplane is around $0.01 (\pm 0.007) c_s$ while the turbulent velocity increases until reaching a quasi plateau with $0.06 (\pm 0.04) c_s$. We note that, dependent on the optical thickness of the line and the local abundance of the molecule, the observations trace different regions inside the disk. The pure vertical component of the turbulent velocity lies slightly below, see dashed line Fig.~\ref{fig:vrms_point}. We also checked for the radial dependence and determined the profiles at 30 and 70 AU. We found that the turbulent velocity slightly increases with radius. At 30 AU the turbulent velocities were reduced by around 30 \% while at 70 AU they were increased by around 50 \% compared to the values at 50 AU.

By carefully investigating for CO, CN and CS observations of the disk system TW Hya, \citet{tea16} found an upper limit of the turbulence causing line broadening, consistent with $v_{turb} \le 0.2 c_s$. Another ALMA observation of the disk HD 163296 in different CO lines by \citet{fla15} supports the general finding of relatively low turbulent velocities in the disk. They found an upper limit of around $v_{turb} \le 0.04 c_s$ for optical thin tracers which are sensitive to the turbulence at the midplane \citep{fla17}, like in the case for DCO+. 

Fig.~\ref{fig:vrms_point}, left panel, shows that the results of our radiation hydrodynamic simulations are consistent with the observations of \citet{tea16} and \citet{fla15} which support the idea of a disk with a relative low turbulent velocity. The midplane and the upper layers lie within the upper limits present by \citet{tea16}. For this comparison, we include our previous magnetized disk simulations presented in \citet{flo15}. In Fig.~\ref{fig:vrms_point}, right panel, we re-analyze the dataset published by \citet{flo15} using model \texttt{D2G\_e-2}, taking the same spatial average and position as described above. In the magnetized model, the turbulent velocities are substantially increased, in the midplane region as well as in the low density upper layers. Starting at a height above  $Z/R>0.25$, which corresponds to $Z>12.5$ AU, the turbulent velocity of the vertical component is still consistent with the proposed upper limit by \citet{tea16} found for TW Hya. 
We note that ambipolar diffusion might reduce the turbulent activity especially in the upper layers of magnetized disks as shown by \citet{sim15b} and so reduce the turbulent broadening of molecular line emission \citep{sim15a}. The details of the region of emission depend mainly on the observed molecule. For TW Hya, \citet{tea16} found the emission from $^{12}$CO to occur at a height $Z/R \sim 0.3-0.4$, from $^{13}$CO at a height $Z/R \sim 0.2$ while the emission from the less optically-thick isotopologue C$^{18}$O traced a height $Z/R \lesssim 0.2$. Observations and models of HD 142527 predict the CO line to originate from a height of about 1 AU at a radius of 100 AU \citep{per15}, which would basically trace the midplane turbulence. Detailed modeling using line radiation transfer and chemistry will be necessary to further constraint the disk regions responsible for the line emission. 

We summarize that the upper limits on the turbulent velocity found in the system TW Hya \citep{tea16} and HD 163296 \citep{fla15} are both consistent with the disk midplane turbulent velocities by magnetized and non-magnetized disk models. 

The turbulent velocity found in the disk atmosphere in magnetized models with Ohmic dissipation only, are to large to be consistent with observations of turbulent line broadening. Further inclusion of ambipolar diffusion could reduce the turbulence to levels consistent with observations. The kinematics of the disk atmosphere is traced by optical thick tracers like $^{12}CO$ or $^{13}CO$. Here, \citet{fla17} found values for HD 163296 which would be more consistent with the non-magnetized model M3, while the predictions for TW Hya \citep{tea16} would compare well with results from both our magnetized and non-magnetized models. 
%
%

\begin{figure*}[ht]
  \resizebox{0.49\hsize}{!}{\includegraphics{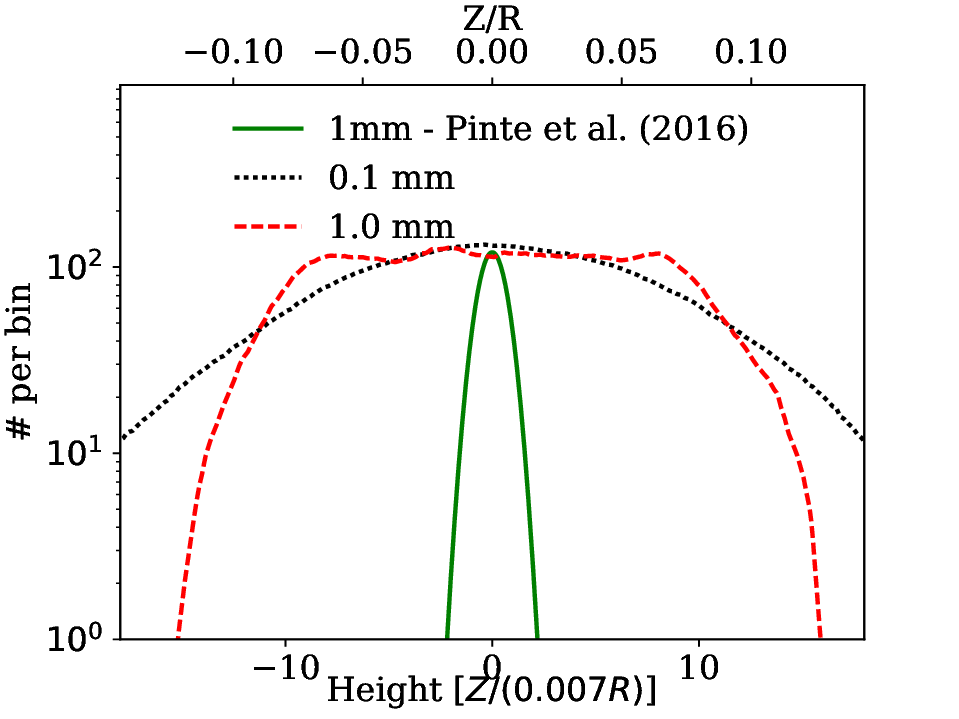}}
  \resizebox{0.49\hsize}{!}{\includegraphics{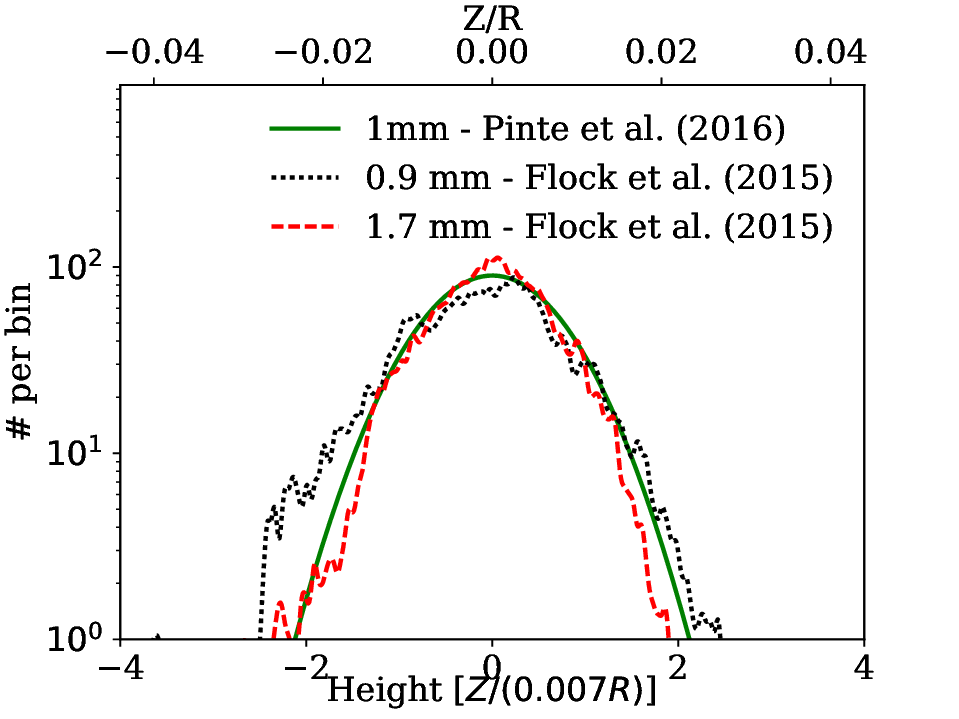}}

\caption{Time and space averaged vertical distribution of two grain sizes. Left: Grain dataset for the model \texttt{M3} taken at 30 AU. Right: Grain distribution for the magnetized disk model by \citet{flo15} and the grain dataset by \citet{rug16} taken at 45 AU.  For both plots, the green line shows the best fit constraints from the systems HL Tau by \citet{pin16} assuming $H/R=0.007$.}  
\label{fig:dust_ver}
\end{figure*}

\subsection{Dust dynamics and scale height} 
\label{sec:dustscale}
\begin{figure*}
  \resizebox{0.5 \hsize}{!}{\includegraphics{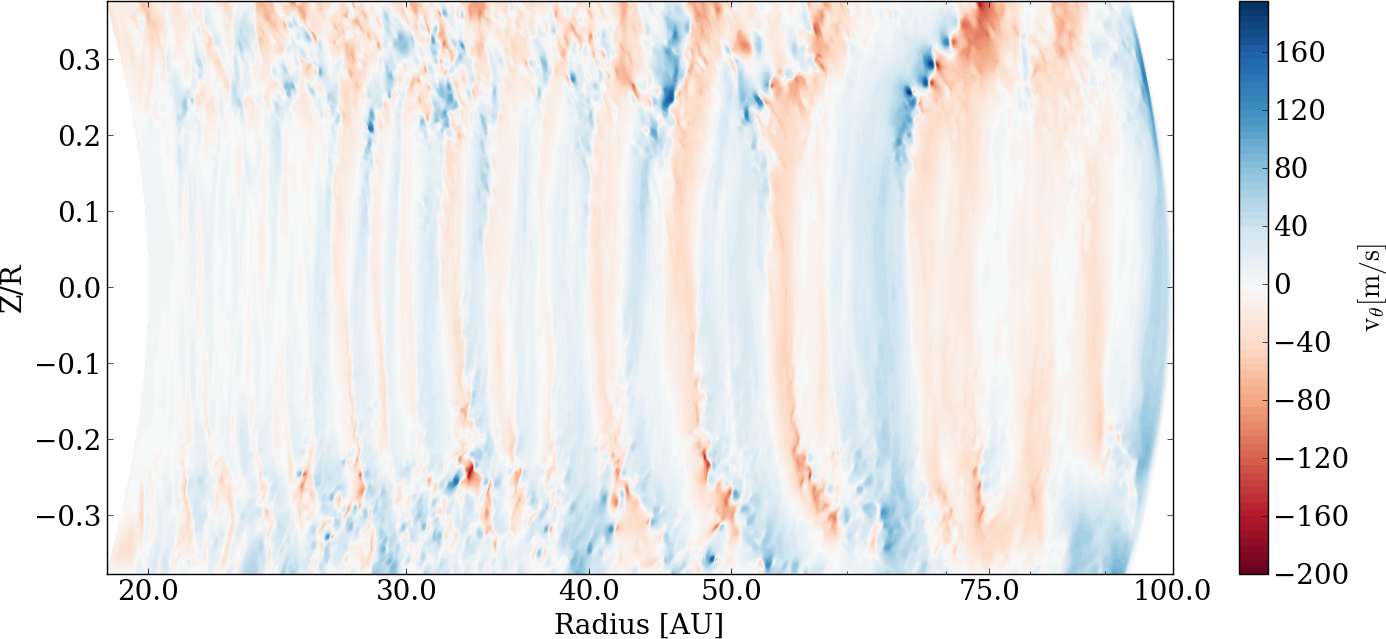}}
  \resizebox{0.5 \hsize}{!}{\includegraphics{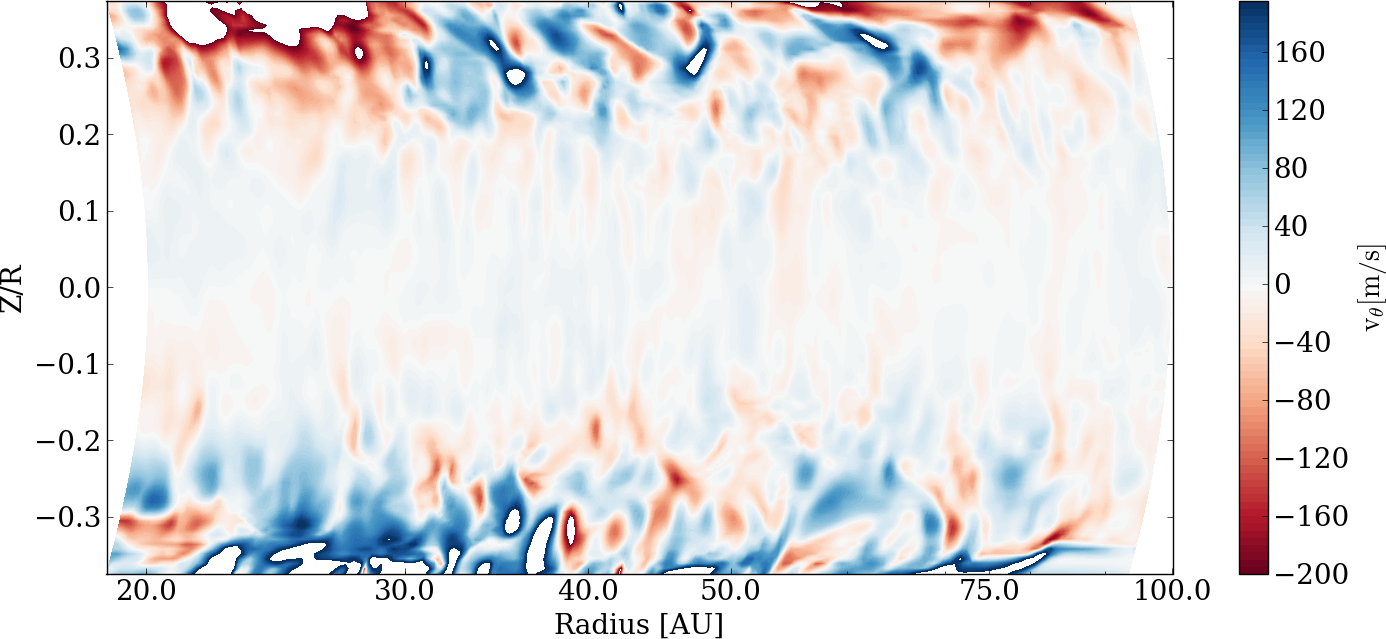}}
\caption{Left: Snapshot of the vertical velocity component in the $R-Z/R$ plane for the model \texttt{M3} taken after 400 inner orbits. Right: Snapshot of the vertical velocity component for the magnetized disk model by \citet{flo15}. Both slices are taken at $\phi=0$.}  
\label{fig:bulkm}
\end{figure*}

In this section we measure the vertical dust distribution obtained in the simulations and compare it with recent observational constraints. In model \texttt{M3} we introduce dust grains with sizes of 100 $\mu$m and 1 mm. The dust grains are spherical and have a solid density of 3 $\rm g\ cm^{-3}$. For this model we consider the Epstein gas drag regime for the particles and the coupling time $t_\text{s}$ becomes
\begin{equation}
t_\text{s} = \cfrac{\rho_\text{d} a}{\rho_\text{g} c_\text{s}},
\end{equation}
with the solid density $\rm \rho_d$, the grain size $a$ the gas density $\rm \rho_g$ and the sound speed $\rm c_s$. For example, grains 1 mm in size have Stokes number $\rm St=t_s \Omega_K=0.053$ in the midplane at 50~AU and so are only moderately coupled to the gas. We note that the stopping time for both particles sizes is well resolved. The particle motion is solved using a Lagrangian method based on the equation of motion in the spherical coordinates system.  For more details on the particle motion routine we refer to the Appendix section by \citet{rug16}. We place a total of one million particles equally distributed over two sizes, and uniformly distributed in the $r-\phi$ plane at the midplane. The initial velocity of the particles is set to match the local Keplerian velocity. The dust grains are distributed at the midplane between 25 and 95 AU. The particles are introduced at a time when the VSI turbulence is already in a steady state after 400 inner orbits. 
To measure the vertical distribution, we consider all particles in the radial bin between 25 and 35 AU. We sample the grains by creating a high resolution grid with 1000 cells in the vertical direction. This grid vertical extent is arbitrarily chosen to include all the grains for the given radial bin. The distribution is averaged for multiple time snapshots after they have relaxed to a new state, see Appendix B. The results are summarized for the two grain sizes in Fig.~\ref{fig:dust_ver}, left panel. The 100 micron grains present a vertical distribution similar to the gas. The 1 mm grains are mixed less efficiently. Their profile shows a quasi-plateau with a dropoff above one gas scale height. This extent of the plateau matches the extent of the vertical bulk motions created by the VSI which appear within one scale height above and below the disk midplane. These large scale vertical oscillations move the bulk of the 1 mm grains up and down. These vertical motions are illustrated in Fig.~\ref{fig:bulkm}. Fig.~\ref{fig:bulkm}, left panel, shows characteristic vertical motions of the disk bulk material of around 50 m/s at the disk midplane for model \texttt{M3}. Those motions easily overshoot the midplane region which emphasizes the need of fully stratified models of both hemispheres. In contrast, the corresponding magnetized model by \citet{flo15} shows turbulent motions with smaller amplitude at the midplane compared to the model \texttt{M3}. In Appendix C we present snapshots of the grains in the R-Z/R plane to show the characteristic vertical motions. We also note that for the given time average only a small number of particles are leaving and entering the radial region used to undertake the average due to radial drift, and therefore this should not affect the results.  

In the following we present and compare the results with our previous magnetized disk models by \citet{flo15} and \citet{rug16}. 
These models share the same stellar and disk parameters as used for this work. In addition to the hydrodynamics those models solve for the magneto-hydrodynamics by considering an initial vertical magnetic field and a 2D $r - \theta $ profile of Ohmic resistivity. Magnetic forces redistribute gas to form a local pressure maximum at the edge of the dead zone, where Ohmic diffusion shuts off magneto-rotational turbulence.  Gas at the pressure maximum orbits exactly at Keplerian speed, halting the headwind-induced inward drift of particles marginally coupled to the outer sub-Keplerian gas by drag.  In addition, the gas within the dead zone, being magneto-rotationally stable, is only weakly turbulent. For these MHD models we analyze the vertical distributions of the grains nearest in size to those we modeled here. These are 0.9 and 1.7 mm in radius, and were trapped between 40 and 50 AU. The results are summarized in Fig.~\ref{fig:dust_ver}, right panel. The particle vertical distribution fits very well with the best-fit model by \citet{pin16}, which determined the vertical scale height for mm size grains in HL Tau to $\rm H_{dust}/R = 0.007$ at 100 AU. The vertical distribution of the 1.7 mm and 0.9 mm grains which are trapped inside the pressure bump lie perfectly in between the best fit value by \citet{pin16}. 

We conclude that the vertical mixing of grains by the vertical shear instability is fundamentally different to the mixing induced by MHD turbulence in the upper layers of protoplanetary disks. 
{The modes that are excited in a VSI-unstable disk have very little structure in the vertical direction compared to the radial direction (i.e. $\rm k_Z/k_R << 1$, where $k_Z$ and $k_R$ are the vertical and radial wave numbers) leading to the efficient lofting of grains away from the midplane and a large effective diffusion coefficient for the vertical mixing of grains.} 
For HL Tau the vertical scale height of 1 mm dust grains was fitted to be small and in the order of $\rm H/R = 0.007$. Our models of grains embedded in a VSI turbulent disk reveal a much larger scale height, close to the gas disk scale height. For HL Tau and HD 163296 such a high dust scale height can be ruled out as otherwise the gaps would not become visible \citep{pin16,ise16}. In contrast, our previous global MHD simulations of a T Tauri system present a dust scale height of mm size grains of $\rm H/R = 0.007$. Those grains are trapped in a ring and they experience relatively low levels of turbulent mixing due to the low ionization degree and hence poor gas-magnetic coupling. {In summary, it appears that the small turbulent velocities inferred for the outer regions of protoplanetary disks from molecular line observations, combined with the high levels of vertical settling inferred for mm-grains, are best explained using a disk model in which non-ideal effects allow low levels of MHD turbulence to be present, but with sufficient magnetic coupling that the VSI is largely suppressed.}

\subsection{Dust continuum observations} 
\label{sec:dustcont}

\begin{figure*}
  \resizebox{\hsize}{!}{\includegraphics{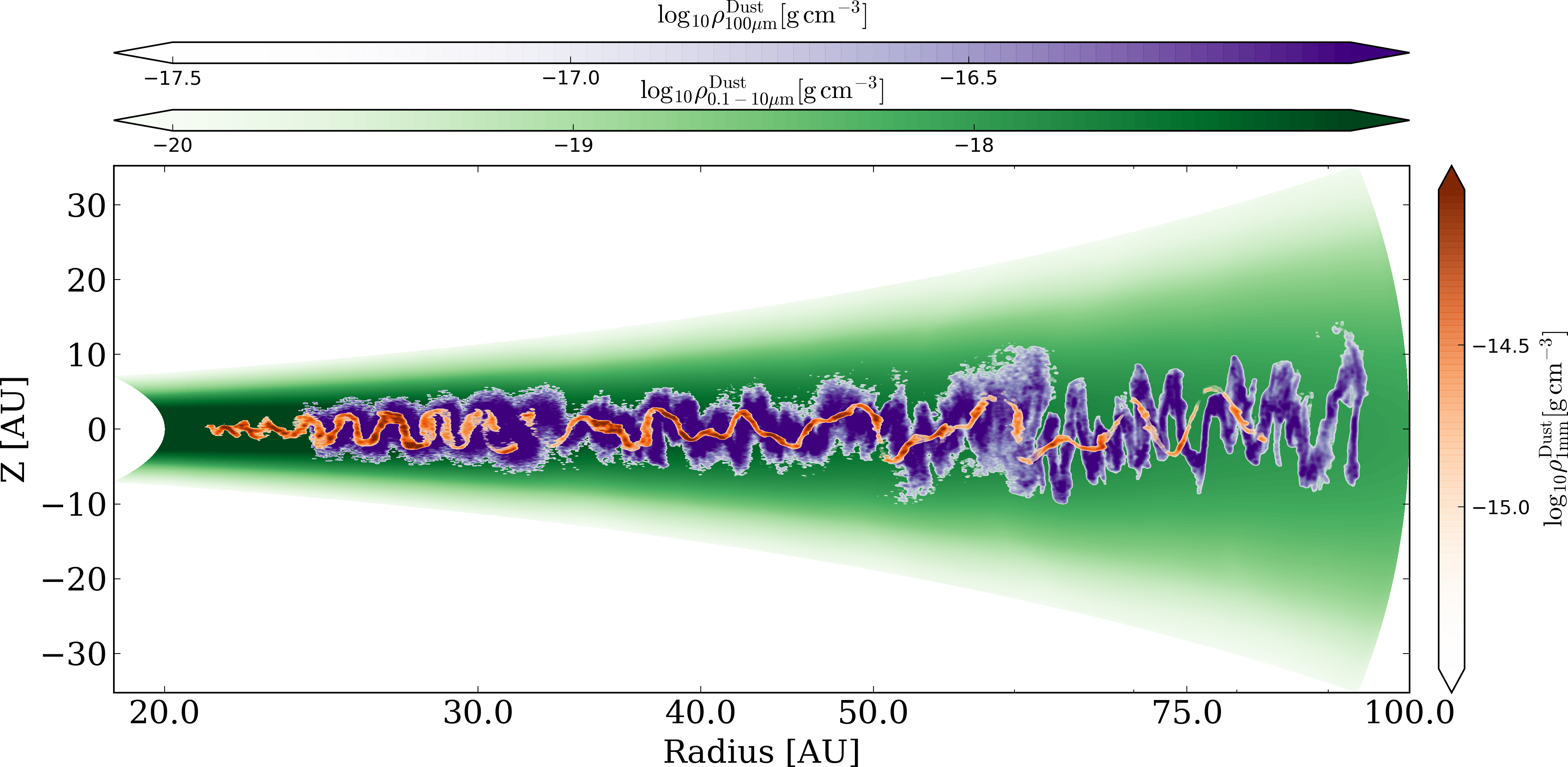}}
\caption{Azimuthally averaged dust density for 3 representative dust grain bins in the $R-Z/R$ plane for the model \texttt{M3} taken after 470 inner orbits (70 orbits of dust evolution). }  
\label{fig:dustdens}
\end{figure*}

\begin{figure}
  \resizebox{\hsize}{!}{\includegraphics{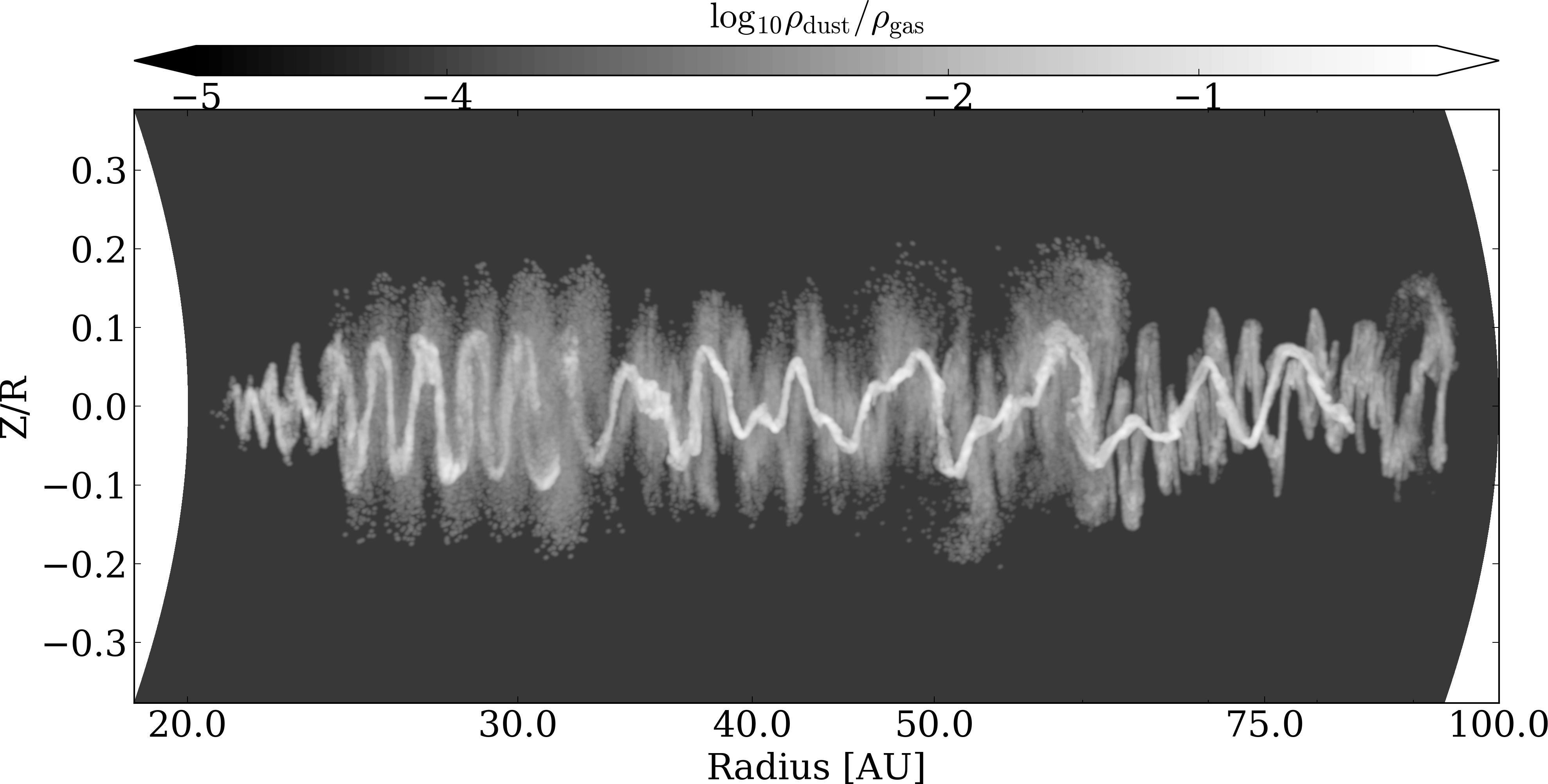}}
\caption{Azimuthally averaged dust to gas mass ratio in the $R-Z/R$ plane for the model \texttt{M3} taken after 470 inner orbits (70 orbits of dust evolution).} 
\label{fig:d2g}
\end{figure}

In the following, we perform Monte-Carlo radiation transfer post-processing of our simulation results with RADMC-3D \citep{dul12}.  
We normalize the dust densities obtained from the radiation-hydrodynamics calculations assuming a global dust-to-gas mass ratio of 1\%, spread over a power-law grain size distribution from 0.1~$\mu$m to 1~mm. The dust density is sampled using 3 dust bins, the small grain bin with a smooth size distribution from $0.1 \mu m$ to $10 \mu m$ and two large grain bins with single sizes of $100 \mu m$ and $1000 \mu m$. To distribute and sample the total dust mass over the bins we follow the prescription presented by \citet{rug16} (Appendix A therein) and using the power-law index of $-3.5$ to determine the size distribution. Fig.~\ref{fig:dustdens} shows a snapshot of the resulting densities of the dust in the three size bins, 0.1-10, 100 and 1000~$\mu$m. The larger grains show stronger settling and the characteristic motions induced by the VSI. The local dust-to-gas mass ratio for the three bins is presented in Fig.~\ref{fig:d2g}. The plot shows that the highest dust concentration and dust density is present for the largest grain size bin. We note that similar structures in the dust density were reported by \citet{lor15}. They presented a new kind of instability caused by the combination of dust settling and the resulting vertical entropy gradient, which they found in gas and dust hydrodynamical simulations.

\begin{figure}
\resizebox{\hsize}{!}{\includegraphics{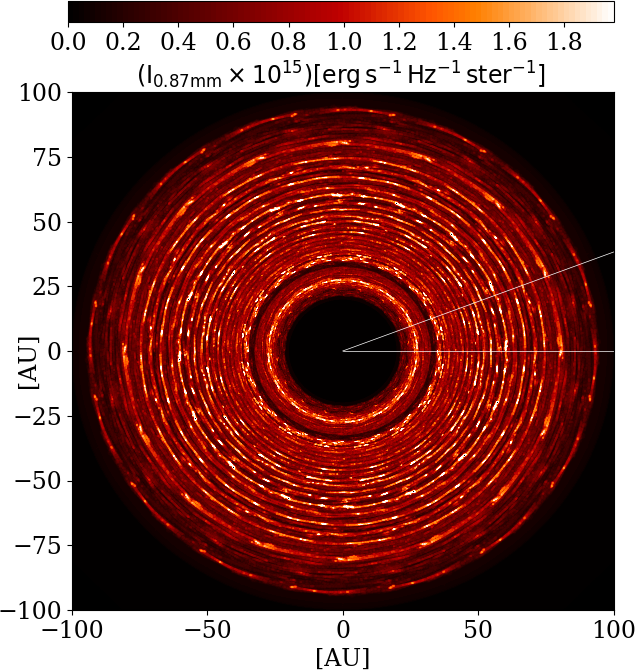}}
\resizebox{\hsize}{!}{\includegraphics{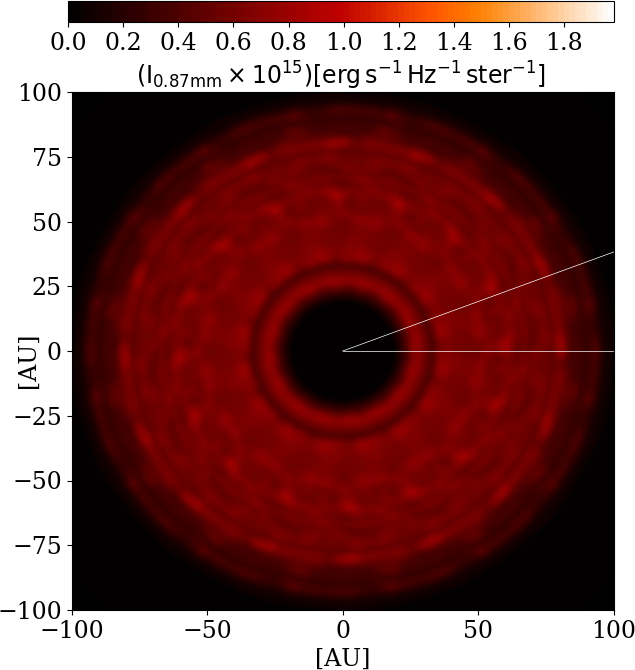}}
\caption{{ Top: Synthetic 0.87-mm image of model M3 viewed face-on. Bottom: Same image convolved with a 2-D Gaussian of half-width 5~AU.  White lines indicate the azimuthal extent of the calculation's domain, $22.5^\circ$.  The rest of the image is generated by applying periodic azimuthal boundary conditions.}}
\label{fig:rt3D}
\end{figure}

We then compute the opacity of the dust in each bin, assuming the grains are made of 62.5 \% silicate and 37.5 \% graphite, using the tool MieX \citep{wol04}. More details on the opacity calculation and the optical properties we use can be found in Appendix C. We pass the results into RADMC-3D to compute synthetic images as in Fig.~\ref{fig:rt3D}, where model M3 is viewed face-on at a wavelength of 0.87~mm. We multiply the intensity by $r^2$ to focus on the appearance of local dust density variations. The strongest features come from radial variations in the density of the largest grains, which are concentrated into rings with azimuthal structure superimposed, similar to \citet{sto16}. The smaller grains are much more smoothly distributed, and contribute little to these features. Note that the grid is carried over from the radiation hydrodynamical calculation, and all features present there are resolved here.  In Fig.~\ref{fig:rt3D}, bottom panel, we show the same synthetic image convolved with a Gaussian filter of width 5~AU ($0.0357^{\prime\prime}$ at 140~pc) to mimic an observation from ALMA. At this resolution, the small-scale features are smoothed out.

The results show that local concentration of the dust material could still be hidden in the large scale structure which is usually observed. This could also be the case for the observation of HL Tau which shows concentric rings \citep{par15}. We note that for our VSI turbulent models we don't observe any long term pressure maximum which would cause a high concentration of particles. Instead, the concentrations observed in the VSI simulations arise because of temporary concentration of grains in short-lived, transient perturbations to the disk vorticity induced by the VSI, as described in \citet{ric16}.

\section{Discussion}
\subsection{Mass accretion rate}
The median rate of mass accretion onto normal classical T Tauri stars is about $10^{-8}$~M$_\odot$~yr$^{-1}$ \citep{naj15,kim16}.  
The Reynolds stresses in our VSI calculations yield much smaller mass flow rates. We estimate the mass accretion rate using the approximation of $\dot{M}= 3 \pi \Sigma \nu$ which is valid in steady state and if $r \gg R_*$. At 100 AU we approximate the mass accretion rate to $\dot{M} \approx 2 \times 10^{-10} M_\sun/yr$ by using $\nu = \alpha H^2 \Omega_K$ with $\alpha=3.8 \times 10^{-5}$ and $\rm H(100\, AU)=14\, AU$ for both of our models. 
 
Recent best-fit estimates from TW Hya by \citet{her08,bri12} determine a mass accretion rate of $10^{-9} M_\sun/yr$. Assuming a constant mass accretion rate over radius, and taking the best fits models of the gas surface density at 100 AU to around $\rm 5 g / cm^2$, $H(100 AU)=9 AU$ and $M_*=0.87 M_\sun$ \citep{menu14, vanb17} we determine a value of around $\alpha^{TW\, Hya}=4 \times 10^{-4}$. Such a value is also preferred in terms of the strength of the dust mixing in their models \citep{vanb17}. 

For the more massive disk around HD 163296, the mass accretion rate were determined recently by \citet{men13} to be $4.5 \times 10^{-7} M_\sun/yr$. Using the same estimate as above with the parameter $\rm 5 g / cm^2$, $H(100 AU)=11 AU$ and $M_*=0.87 M_\sun$ \citep{ise16} yields a value of $\alpha^{TW\, Hya}=7.6 \times 10^{-2}$. However in this system, such a high value can be ruled out due to the good visibility of the gaps \citep{ise16}. 

%
%
%
Even given these uncertainties for the estimates of the gas surface density and the fact that the mass accretion rates vary over radius and time, we think that it is unlikely that the VSI alone can explain the observed accretion rates. At best, given the current results and constraints the VSI could support the mass accretion rate at larger radii where the conditions required for short thermal equilibration timescales are fulfilled. In contrast, magnetized disk models predict a turbulent stress level two orders of magnitude higher at similar distances from the star \citep{sim15b,flo15}.

\subsection{Kinematic constraints by observations and VSI kinematic}
The observations of the close-by young protoplanetary disk TW Hya with ALMA by \citet{tea16} have demonstrated the difficulty of constraining the exact value of the gas turbulence in the disk by means of molecular line observations. There are two main limitations to determine the gas kinematics from the broadened line profile. The first limitation comes from the fact that the thermal broadening and the turbulent broadening are difficult to disentangle. In the work by \citet{fla15,fla17}, and also pointed out by \citet{sim15a} , they are able to disentangle these effects by viewing the molecular line in the spatial domain. However we note that if the thermal and turbulence structure is unknown, the systematic error in determining the turbulent broadening could be as high as the broadening itself. The second limitation comes from the flux calibration error. \citet{tea16} pointed out that the flux calibration error alone limits the accuracy with which the turbulent broadening component can be determined. Improved observations and modeling are needed to further constraints the kinematics.

We also emphasize that the turbulent structure caused by the VSI are clearly different than the turbulence generated by the MRI. While the MRI generates a turbulent structure which would produce a thermal-type line broadening \citep{sim15a} this is not straightforward applicable for the VSI. Especially the large scale motions might produce a broadening type which could be different and depend on the inclination of the disk. Further detailed line transfer models should be done to constrain the detailed effect of the VSI turbulence onto the line shape. Especially spatially resolved observations would allow to obtain radial profiles of the turbulent line broadening to further constrain the underlying mechanism to trigger turbulence in disks. 

\subsection{Dust feedback and VSI activity}
We briefly want to address important implications the results could have regarding the dust motion.
The radiation hydrodynamical simulations of dust and gas show coherent motions of the large grains. They will naturally lead to very low relative velocities between two neighboring large grains. This could suppress fragmentation and lead to much larger grains sizes which would then totally decouple from the gas motion \citep{oku12}. Secondly the local dust density can become quite large for the mm size grains. In the 3D dataset shown in Fig.~\ref{fig:d2g}, we found the maximum dust to gas mass ratio in one single cell to be $2.3$. This substantially increases the collision rate and so decreases the grain growth timescale \citep{bra08,bir12}. We expect the dust drag force to dominate at the midplane region and therefore we expect only little effects on the VSI activity. However the dust drag could affect the dynamics at the midplane region. Future simulations including the dust feedback onto the gas are needed to further constrain these effects.

\section{Summary and conclusion}
We have performed 3D radiation hydrodynamical simulations of protoplanetary disks' outer regions, resolving all three dimensions
with about 70~cells per density scale height.  The star and disk parameters are those of a typical T~Tauri system. The disk absorbs
and reradiates the starlight with opacities computed using Mie theory, and dominated by small dust grains. We consider dust-to-gas mass
ratios of the small grains of $10^{-4}$ and $10^{-3}$, leading to thermal equilibration rapid enough for the vertical shear instability to operate. We also track the movements of 0.1 and 1~millimeter grains embedded in the two model disks.  Our main findings on the gas and dust dynamics are:


\begin{enumerate}
\item The VSI saturates in turbulence that remains in quasi-steady state out to at least 1600~inner orbits (143~outer orbits). The mass-weighted stress-to-pressure ratio is about $4\times 10^{-5}$ at both dust fractions. The turbulent gas motions have RMS Mach number $\sim 1$\% at the midplane, and $\sim 10$\% in the corona.

\item The VSI produces clear vertical bulk motions, which loft the 1-millimeter grains to about one gas scale height from the midplane. This is much higher than the accretion stresses would imply under isotropic turbulent diffusion, thanks to the anisotropy. The characteristic radial extent of the upward- or downward-moving gas parcels is also about one scale height.

\end{enumerate}
Further, we compared our models with observational constraints, and also with results from a magnetized disk model by \citet{flo15}
and \citet{rug16}. The main findings are:
%
\begin{enumerate}
\item Turbulent speeds at the disk midplane in both the magnetized and non-magnetized models are broadly consistent with upper limits on turbulent line broadening measured in the TW~Hya disk at 0.2~times the sound speed $c_s$ \citep{tea16} and in HD~163296 at 0.04$c_s$ \citep{fla15}. The VSI turbulence remains below the upper limit of 0.2$c_s$ over the whole domain, while the magnetized disk models' corona violates this constraint above height $Z=0.2R$. The molecular line observations are sensitive to regions outside 10~AU, and can trace the interior $Z\ll 0.2R$ as in HD~142527 \citep{per15} or corona $Z\sim 0.2R$ as in TW~Hya \citep{tea16}, depending on the disk surface density and the molecule.

\item The mm-sized particles' scale height in the VSI turbulence is too great to be consistent with the well-separated rings of millimeter thermal emission observed in the inclined HL~Tau disk. Such thick rings would overlap one another along the projected minor axis \citep{pin16,ise16}. In contrast, the grains trapped near the dead-zone edge in the magnetized calculation well-match the scale height $H_{\rm dust}=0.007R$ that \citet{pin16} inferred for the mm grains in HL~Tau.

\item Synthetic images made from our global radiation hydrodynamical results show concentric rings much narrower than the gas scale height. When convolved with a nominal telescope beam, the rings appear wider and their amplitude is reduced. The rings are transient, with the grains experiencing neither long term trapping nor net radial concentration.

\item The VSI turbulent Reynolds stress is two order of magnitude less than needed to explain measured rates of accretion onto the stars. Angular momentum is thus probably not extracted purely by VSI.



\end{enumerate}

Overall, these results are consistent with a picture in which magnetic forces drive the accretion flow and suppress the VSI, but the magnetic
activity is modulated by non-ideal MHD effects. Thus, detecting magnetic fields and measuring their strengths would be extremely useful in understanding protostellar disks' dynamics and evolution. Furthermore, future modeling should treat the grains' movements through the gas, because most of the solid material in protostellar disks' outer regions shows significant settling.

\section*{Acknowledgments}
The authors thank Sebastian Wolf and Hubert Klahr for their helpful discussions and comments on the project. We thank Andrea Mignone for supporting and advising us with the newest PLUTO code. We thank Andrea Isella for his comments on the radiation transfer post-processing. Parallel computations have been performed on the zodiac supercomputer at JPL. For this work, Mario Flock received partial funding from the European Research Council under the European Union's Seventh Framework Programme (FP7/2007-2013) / ERC Grant agreement nr. 258729. This research was carried out in part at the Jet Propulsion Laboratory, California Institute of Technology, under a contract with the National Aeronautics and Space Administration and with the support of the NASA Exoplanet Research program via grant 14\-XRP14\_2\-0153. Richard Nelson acknowledges support from STFC through the grants ST/P000592/1 and ST/M001202/1. Gesa H.-M. Bertrang acknowledges financial support from CONICYT through FONDECYT grant 3170657 as well as by the Millennium Science Initiative (Chilean  Ministry  of  Economy),  through grant Nucleus RC13007. Wladimir Lyra acknowledges support of Space Telescope Science Institute through grant HST-AR-14572 and the NASA Exoplanet Research Program through grant 16-XRP16 20065. This research was supported in part by the National Science Foundation under Grant No. NSF PHY-1125915. Copyright 2017 California Institute of Technology. Government sponsorship acknowledged. 

\bibliographystyle{apj}
\bibliography{VSI}

\appendix 

\section{{\bf A} Longterm turbulent evolution and calculation of the Reynolds stress}
\begin{figure}
  \resizebox{0.49 \hsize}{!}{\includegraphics{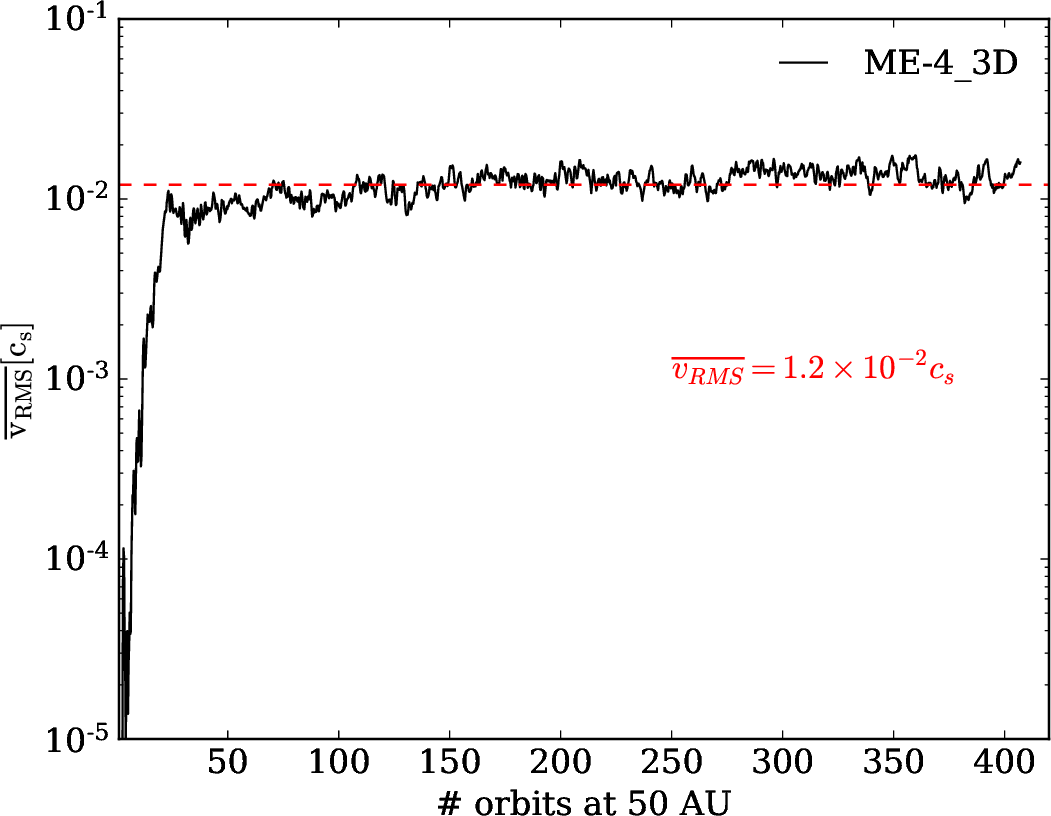}}
  \resizebox{0.49 \hsize}{!}{\includegraphics{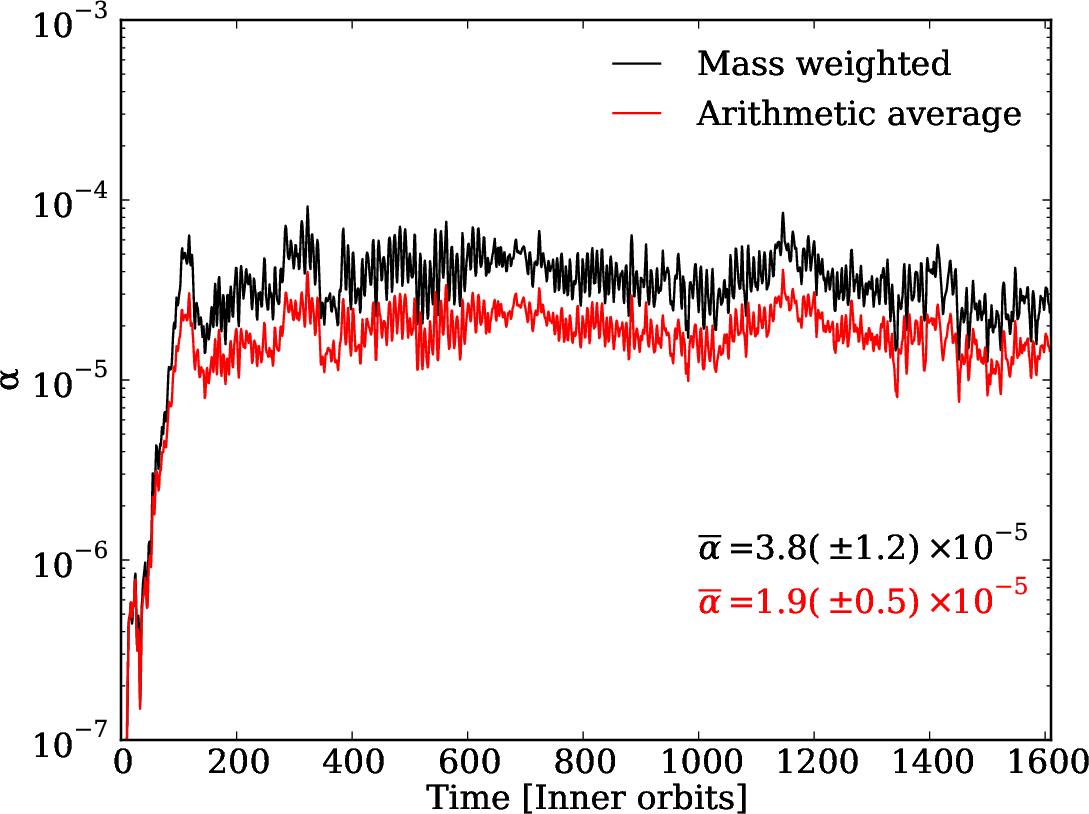}}
\caption{Left: Longterm time evolution of the mass weighted turbulent velocity over radius for model \texttt{M4}. Time averaged values are annotated using a red dashed line and labeling. Right: Corresponding time evolution of the stress-to-pressure ratio $\alpha$, using the approach of mass weighted values (black) and using an arithmetic average (red).} 
\label{fig:ref}
\end{figure}

The longterm turbulent evolution of the disk was calculated for model \texttt{M4}. To capture the growth of modes with longer growth times we run the model for 1610 inner orbits (143 outer orbits). The detailed growth of the turbulent velocities is shown in Fig.~\ref{fig:ref}, left. 
At 30 local orbits the simulation reaches a clear steady state showing mass weighted turbulent velocities at around one percent of the sound speed. The longterm evolution of the stress-to-pressure ratio is shown in Fig.~\ref{fig:ref}, right. The simulation shows a clear steady state. The turbulent velocity and the stress-to-pressure ratio are mass weighted. 

To compare our results on the calculation of the Reynolds stress we present here another approach to calculate the time and space averaged stress-to-pressure ratio. For this second approach we follow the calculation described by \citet{nel13}. For this, we calculate the Reynolds stress $T_{r \phi} (r,\theta)=\rho \delta v_r \delta v_\phi$ in the 2D plane ($r$,$\theta$) by averaging along azimuth. We then define a density weighted, azimuthally and vertically averaged, mean pressure $\overline{P}(r)$ at reach radius r, which allows then to express the stress-to-pressure ratio as $\alpha(r,\theta)= T_{r \phi} (r,\theta) / \overline{P}(r)$. In Fig.~\ref{fig:ref}, right, we compare the results. The arithmetic average results into a slightly smaller value of the stress-to-pressure ratio. This can be explained as the value of $\alpha$ slightly increases with radius. As there is more mass stored at the outer radii, the mass weighted value is slighlty higher. Overall, the value of the stress-to-pressure ratio proofed to be small and slightly lower compared to what was found in previous global isothermal \citep{nel13} and recent global radiation hydrodynamic models \citep{sto16}, and probably depend in detail on the thermodynamic and opacity models adopted. Certainly, further high-resolution studies of the VSI are needed to constrain the level of turbulence, especially due to the fact that the fastest growing modes have very short wavelengths and usually are limited by the grid resolution \citep{bar15}.



 

\section{{\bf B} Vertical bulk motions by the VSI}
\begin{figure}
\resizebox{\hsize}{!}{\includegraphics{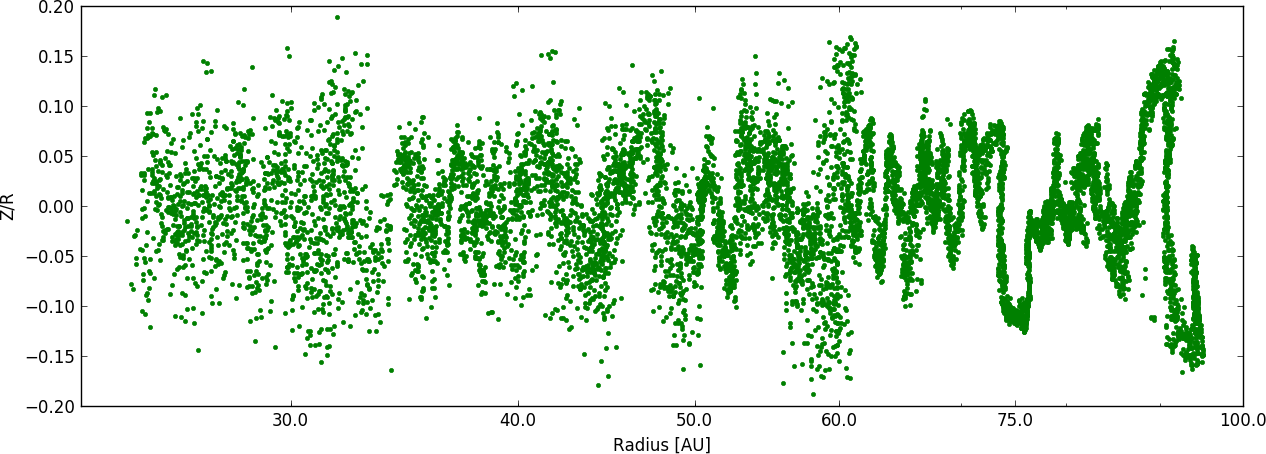}}
\resizebox{\hsize}{!}{\includegraphics{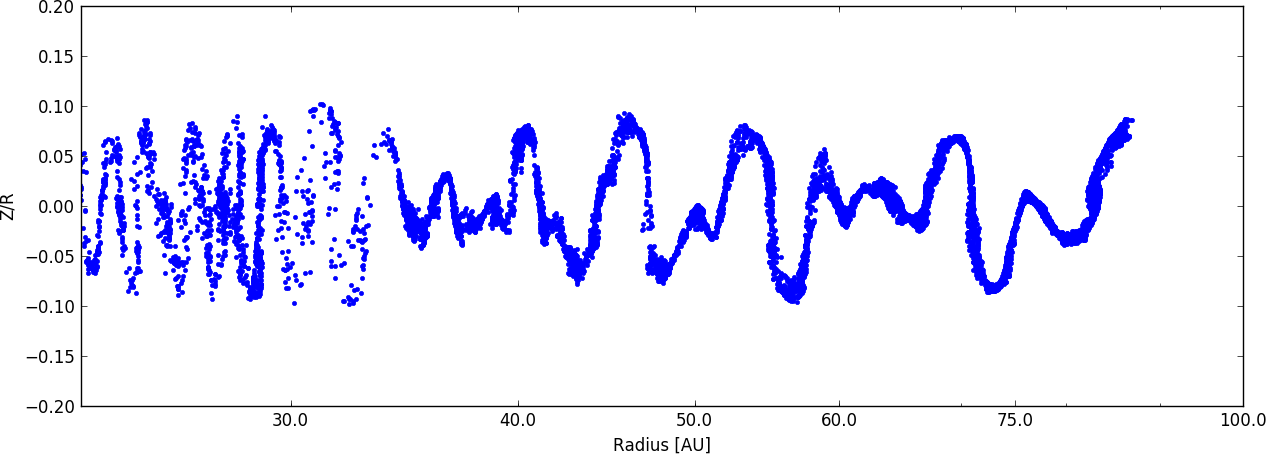}}
\caption{Snapshot of the grain distribution in the $\rm R-Z/R$ plane for the 0.1 mm grains (top) and for the 1 mm grains (bottom) after 50 inner orbits of evolution of model \texttt{M3}. Every fiftieth particle from the total dataset is shown.}
\label{fig:par_mov}
\end{figure}

In this appendix section we present snapshots of the grain distribution after 50 inner orbits of evolution. Fig.~\ref{fig:par_mov} shows the snapshot of the evolution of 50 inner orbits. For the two grain sizes, the vertical bulk motions are able to lift the grains $\pm$ one scale height upwards and downwards the midplane. These motions are responsible for the effective large scale height described in Section~\ref{sec:dustscale}. 

\begin{figure}
\resizebox{0.5\hsize}{!}{\includegraphics{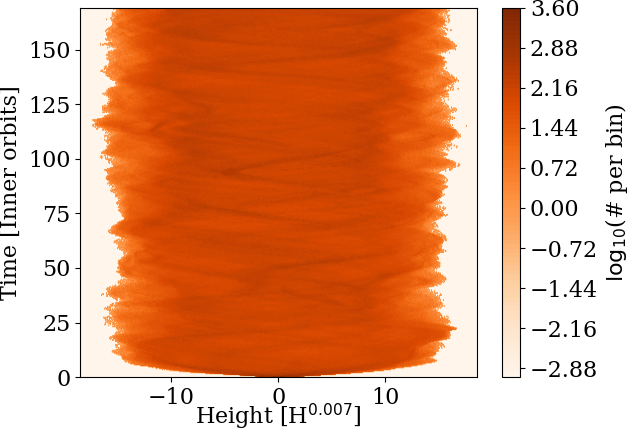}}
\caption{Time evolution of the 1 mm grains vertical distribution of model \texttt{M3}.}
\label{fig:par_vsi_vert}
\end{figure}

In the following we will check how fast the millimeter grains reach an equilibrium state. Fig.~\ref{fig:par_vsi_vert} shows the vertical distribution of mm grains over time in the radial extent from 25 to 35 AU. The particles quickly reach into an equilibrium state. To determine the vertical profile shown in Fig.~\ref{fig:dust_ver}, we have taken a time average starting after 20 inner orbits. 

\section{{\bf C} Calculation of the opacity and RT post-processing}
To determine the opacity we use the tool MieX \citep{wol04} using the optical data by \citet{wei01}. The dust material is composed of $62.5\, \%$ silicate, 25 \% of perpendicular graphite and 12.5 \% of parallel graphite. We determine three characteristic dust bins. The first bin uses minimum and maximum grain sizes of $a_{min}=0.1 \mu m$ and $a_{max}=10 \mu m$ with the size distribution exponent of $-3.5$. The second and third size bin uses single dust grain sizes of 0.1 mm and 1 mm. The wavelength grid is the same to determine the opacity and to calculate the RT modeling. From MieX we obtain the scattering and absorption efficiency factors $Q_{sca}$ and $Q_{abs}$. Together with the corresponding cross section $C=Q \pi a_{eff}^2$ we can determine the opacity per gram of dust using
\begin{equation}
\kappa_\lambda = \frac{Q^{3/2}}{C^{1/2}} \frac{\sqrt{\pi}} {\rho_d} \frac{3}{4},
\end{equation}
for each wavelenght. For the RADMC-3D setup run we use $10^9$ photon packages. We apply the same grid in RADMC-3D as it was used for the global simulation run. We use 3 individual dust grain bins and we determine the particle density from the corresponding particle distribution.   



\end{document}